%% file: main.tex
\documentclass[conference]{IEEEtran}
\IEEEoverridecommandlockouts
\usepackage{cite}
\usepackage{amsmath,amssymb,amsfonts}
\usepackage{graphicx}
\usepackage{textcomp}
\usepackage{xcolor}
\def\BibTeX{{\rm B\kern-.05em{\sc i\kern-.025em b}\kern-.08em
    T\kern-.1667em\lower.7ex\hbox{E}\kern-.125emX}}

\usepackage{caption}
\usepackage{url} 
\usepackage{comment}
\usepackage{listings}
\usepackage{algorithm,algorithmic}
\usepackage{pifont}
\usepackage{subcaption}

\begin{document}
\title{SoK: Liquid Staking Tokens (LSTs)\\and Emerging Trends in Restaking} 


\author{Krzysztof Gogol \\ \small University of Zurich \\ \small gogol@ifi.uzh.ch
    \and Yaron	Velner \\ \small Risk DAO \\ \small  yaron@riskdao.org
    \and Benjamin Kraner \\ \small University of Zurich
    \and Claudio Tessone \\ \small University of Zurich \\ \small UZH Blockchain Center
}

\maketitle

\begin{abstract}\input{sections/00Abstract}
\end{abstract}


\input{sections/01Introduction}
\input{sections/02Background}
\input{sections/03LiquidStaking}
\input{sections/05LST_Protocols}

\input{sections/04Restaking}
\input{sections/06LRT_Protocols}

\input{sections/08Discussion}
\input{sections/09Conclusions}

\bibliographystyle{IEEEtran}
\bibliography{main}

\vspace{12pt}

\appendix
\input{sections/98EthereumSecurity}

\end{document}

%% file: sections/00Abstract.tex
Liquid staking and restaking represent recent innovations in Decentralized Finance (DeFi) that garnered user interest and capital. Liquid Staking Tokens (LSTs), tokenized representations of staked tokens on Proof-of-Stake (PoS) blockchains, are the leading staking method. LSTs offer users the ability to earn staking rewards while maintaining liquidity, enabling seamless integration into DeFi protocols and free tradeability. Restaking builds upon this concept by allowing staked tokens, LSTs or native Bitcoin tokens to secure additional protocols and PoS chains for supplementary rewards. 
Liquid Restaking Tokens (LRTs) unlock liquidity of restaked assets.
This Systematization of Knowledge (SoK) establishes a comprehensive framework for the technical and economic models of liquid staking protocols. Using this framework, we systematically compare protocols mechanics, including node operator selection, staking reward distribution, and slashing. Our empirical analysis of token performance reveals that protocol design and market dynamics impact token market value. 
We further present the recent developments in restaking and discuss associated risks and security implications. Lastly, we review the emerging literature on liquid staking and restaking.

%% file: sections/01Introduction.tex
\section{Introduction}
\label{sec:introduction}

Ethereum~\cite{Buterin2014Ethereum}, the largest Proof-of-Stake (PoS) blockchain by market capitalization and staking volume~\cite{2023StakingRewards}, transitioned from Proof-of-Work (PoW) to PoS through "The Merge" upgrade in 2022-2023~\cite{2023EthereumRoadmap}. This transition reduced Ethereum's energy consumption by 99.99\%~\cite{2023EthereumExpenditure} and enabled users to stake Ether (ETH) to participate in transaction validation, network maintenance, and earn staking rewards. 

While staking offers consistent rewards~\cite{Carre2022Security}, it locks tokens, limiting liquidity. Users can stake tokens in PoS blockchain through various means: running their own validator nodes, using centralized exchanges, staking platforms, investment products like ETFs, or via \emph{Liquid Staking Tokens} (LSTs)—tokenized representations of staked token. LSTs have become the dominant staking method for ETH, representing 37\% of total staked ETH, surpassing centralized exchanges (28\%) and staking pools (15\%)~\cite{2023DuneETHStaking} at the end of 2023.

LSTs enhance flexibility by allowing tokens to accumulate staking rewards while remaining tradable and usable in decentralized finance (DeFi) protocols~\cite{Gogol2023SoK:Risks}. Key benefits include: (i) enabling trading and transfers, (ii) serving as dominant collateral in DeFi lending~\cite{2023MessariQ123Report}, (iii) supporting the minting of decentralized stablecoins, (iv) facilitating liquidity provision in automated market makers (AMMs), and (v) multiplying staking rewards by leveraging LSTs. Moreover, some liquid staking protocols (LSPs) lower the validator threshold, allowing users to operate validators with as little as 4 or 8 ETH instead of the standard 32 ETH.

Consequently, liquid staking has emerged as the largest DeFi category, with \$58 billion in Total Value Locked (TVL). Lido~\cite{2020Lido:Whitepaper}, the first protocol offering LSTs, became the largest DeFi protocol by TVL~\cite{2022DeFiLlama}. However, Lido’s 32\% share of the total staked ETH in 2023 raised concerns about the risks of centralization of the Ethereum network.
Protocols implementing liquid staking differ in governance and design, including validator operations, reward distribution, and slashing mechanisms. For instance, centralized models like Lido, which relies on 30 node operators, contrast with decentralized approaches such as Rocket Pool, which boasts over 3,000 operators~\cite{2023DuneLSDDeposit}. Distributed Validator Technology (DVT)~\cite{2023EthereumDVT} further enhances security by distributing validator keys across multiple operators, mitigating centralization risks.

\emph{Restaking}, introduced by EigenLayer~\cite{2023EigenLayer:Whitepaper}, builds upon staking by enabling staked tokens or LSTs to secure additional applications, known as \emph{Actively Validated Services} (AVSs), while earning extra rewards. Typical AVS use cases include data availability layers, bridges, and oracle services. With TVL exceeding \$15 billion, EigenLayer is the third largest DeFi protocol~\cite{2022DeFiLlama}.

Recently, restaking was extended to \emph{cross-chain security}, in which tokens staked on one PoS chain (the provider chain) are restaked to secure another PoS chain (the consumer chain). This approach, also called \emph{mesh security}, was first proposed for the Cosmos ecosystem~\cite{2022MeshSecurity,2023MeshSecurity}.
Another form of cross-chain security is \emph{Bitcoin staking}, introduced by the Babylon protocol~\cite{Babylon2023BitcoinStaking}. Similarly to Ethereum restaking and Cosmos mesh security, Bitcoin staking allows BTC holders to secure other protocols, including staking on PoS chains. Notably, cross-chain restaking, including Bitcoin staking, provides slashable security guarantees without bridging tokens to the consumer PoS chain. Babylon Protocol is currently the second largest restaking protocol with TVL of more than \$5 billion~\cite{2022DeFiLlama}.

\emph{Liquid Restaking Tokens} (LRTs) extend the concept of restaking, unlocking liquidity for tokens used as collateral in AVS or restaked at consumer PoS chains. Both restaking and liquid restaking have grown in popularity, with TVL reaching \$25 billion and \$15 billion, respectively.
Although LSTs and LRTs improve user rewards, they also introduce additional risks. Furthermore, as pegged tokens, their market values oscillate around the total value of staked or restaked tokens and are vulnerable to temporary depegging, a risk empirically examined in this study.

\subsection*{Contribution}

This work presents a systematic study of liquid staking protocols, including emerging solutions for Ethereum restaking, Bitcoin staking and cross-chain security. It introduces a taxonomy of liquid staking protocols and tokens and empirically examines their impact on token value. 
The analyses in this study are based on whitepapers, technical documentation, and GitHub repositories of major liquid staking and restaking protocols. The focus of the liquid staking analysis is on Ethereum, which accounts for 91\% of the liquid staking market. The restaking study is centered on EigenLayer and Babylon protocols, which together represent over 82\% of the restaking market. For context and comparison, we also mention complementary approaches such as Mesh Security on Cosmos.
This work makes the following key contributions:

\paragraph{Generalizing Mechanisms and Taxonomy} We generalize the mechanisms and economics of liquid staking, protocols and tokens. This includes a summary of their key properties, which establishes a taxonomy for node operations, reward distribution, and token models. We discuss and systematize the risks related to liquid staking from the perspective of token holders, DeFi users, and underlying PoS chains. 
    
\paragraph{Positioning Within DeFi} We position liquid staking tokens within the broader taxonomy of Decentralized Finance (DeFi) and pegged tokens, examining their relationships and interactions with other DeFi protocols. The analysis includes limitations and empirical testing of LST performance in tracking staking rates, linking performance to architectural decisions. LSTs operate similarly to stablecoins and wrapped tokens in maintaining their peg. 

\paragraph{Pegged Token Empirical Analysis} Our analysis reveals that while LST market prices track staking rewards with comparable accuracy, temporary de-pegs occur during extreme market events (e.g., FTX insolvency or Terra/Luna crash). Decentralized LSTs experience upward de-pegs when insufficient node operators are available to run validators. Importantly, not all price disparities between market prices (at DEXs) and protocol prices (based on staked reserves) lead to arbitrage opportunities due to the underlying architecture decisions.

\paragraph{Literature Review and Future Directions} We provide a review of the literature summarizing current research priorities on liquid staking and restaking.

\subsection*{Paper Organization}

Section \ref{sec:background} provides a brief overview of the Proof-of-Stake (PoS) mechanism, with essential background to understand how stake rates are determined in Ethereum.
Section \ref{sec:liquidstaking} establishes a taxonomy of Liquid Staking Tokens (LSTs), and introduces Distributed Validator Technology (DVT) as a solution to enhance protocol resilience.
In Section \ref{sec:empirical}, the paper empirically analyzes the major LSTs protocols.
Section \ref{sec:restaking} presents the recent related developments in the area of restaking, including Bitcoin staking and cross-chain restaking and Liquid Restaking Tokens (LRTs).
Section \ref{sec:discussion} provides a discussion of the empirical results and associated risks. 
Section \ref{sec:relatedwork} summarizes related work, while Section \ref{sec:conclusions} summarized and provides the conclusions of this study.
Appendix \ref{sec:ETHSecurity} evaluates the efficacy of DVT in mitigating security vulnerabilities in PoS consensus arising from liquid staking.

%% file: sections/02Background.tex
\section{Background}
    \label{sec:background}
This section provides an overview of the Proof-of-Stake (PoS) consensus mechanism, the calculation of staking rewards, and Ethereum's transition to PoS through the upgrade process known as \emph{The Merge}.

\subsection{Proof-of-Stake Consensus}
    \label{sec:PoW_PoS}

Proof-of-Work (PoW) and Proof-of-Stake (PoS) are the two primary consensus mechanisms used in blockchain networks to validate transactions and achieve agreement on the ledger's state.
In PoW, introduced by Bitcoin \cite{Nakamoto2008Bitcoin:System}, miners compete to solve complex mathematical puzzles, expending significant computational power and energy. The first miner to solve the puzzle is rewarded with newly minted cryptocurrency and transaction fees, and their validated block is added to the blockchain. The security of PoW relies on the assumption that the majority of computational power in the network is controlled by honest participants, preventing double-spending and tampering with the blockchain's history.

In contrast, PoS offers an energy-efficient alternative. Validators in PoS blockchains create new blocks, attest proposed blocks and secure the network. They are required to lock a certain amount of native tokens and "stake" them as collateral, currently 32 ETH in the Ehtereum blockchain. Unlike miners in PoW, validators in PoS are guaranteed rewards over time, provided they fulfill their duties. Validators are penalized for inactivity, missed proposals, or malicious behavior through a process called \emph{slashing}. 

\subsection{Ethereum Staking}

Staking Ether (ETH) involves participating in Ethereum's PoS consensus mechanism by locking a minimum of 32 ETH as collateral in a validator. During staking, validators deposit ETH into a smart contract for a specified period, actively engaging in block validation and consensus-related tasks. This process ensures the security of the Ethereum network, as the PoS mechanism assumes that most participants act honestly to protect their staked assets.

\begin{figure*}[!th]
  \centering
    \includegraphics[width=0.85\textwidth]{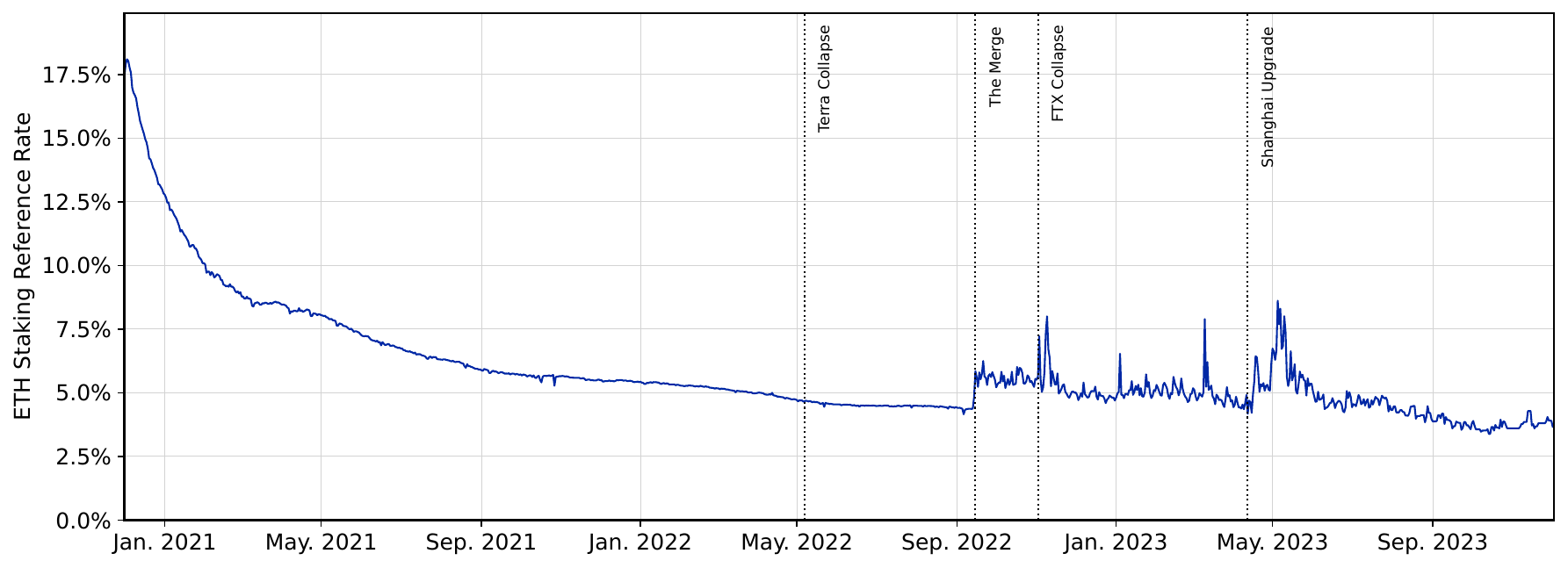}
\caption{Annualized historical staking rewards for Ethereum during its transition from PoW to PoS, concluded in the Shanghai Upgrade (Data source: ETH.STORE).}

    \label{fig:apy}
\end{figure*}

Validators are incentivized to adhere to protocol rules and act in the network's best interest through the promise of rewards. These rewards can be classified into two categories: indigenous rewards, which are intrinsic to Ethereum's PoS design, and exogenous rewards, which arise from external opportunities like Maximal Extractable Value (MEV)~\cite{Heimbach2022MEV}. Figure \ref{fig:apy} illustrates the historical annualized staking rewards on the Ethereum blockchain, showcasing the incentives available to validators for their contributions.

\subsubsection*{Indigenous Staking Rewards}
Indigenous rewards are specific to Ethereum's PoS mechanism and are guaranteed to validators provided they fulfill their assigned duties. These rewards include attestation, block proposal, sync committee participation, slashing rewards, and priority fees \cite{2023RocketPool:Documentation}. The Ethereum PoS protocol calculates these rewards based on key factors such as the participation rate, total staked ETH, epoch length, and network inflation \cite{2023EthereumStaking}.

\paragraph{Participation Rate} Also referred to as the effective balance, this represents the actively staked ETH by a validator. Validators with higher effective balances are more likely to be chosen for block validation and earn rewards.
 \paragraph{Total Staked ETH} The aggregate ETH staked across the network determines a validator’s share of the overall stake and their probability of selection.
\paragraph{Epoch Length} Staking rewards are distributed at the end of epochs, which are time intervals of approximately 6.4 minutes. The duration and frequency of these epochs directly influence reward distribution.
\paragraph{Network Inflation} Staking rewards are influenced by the inflation rate, which dictates the minting of new ETH tokens and their distribution to validators. This rate dynamically adjusts based on network parameters and rules.

\subsubsection*{Exogenous Staking Rewards}
An additional source of revenue for validators is Maximal Extractable Value (MEV). MEV refers to the re-ordering transaction in the block by the validator to generate profit. Various forms of MEV exist e.g. front-running and back-running, and the revenue comes from the arbitrage opportunities at DEX, liquidations at lending protocols, or sandwich attacks \cite{Chitra2022ImprovingRedistribution, Heimbach2022MEV}. Validators participate in MEV auctions and compete with other validators to win the MEV rewards. These rewards are not deterministic and independent of the Ethereum PoS mechanism and, consequently, are referred to as exogenous rewards.

\subsubsection*{Staking and Unstaking Queues}
Ethereum PoS introduces \emph{unstaking queues} for validators to ensure network stability. The length of the activation and exit queues depends on several factors, including the number of validators joining or exiting, the total number of active validators in the network, and the churn limit. 
A maximum of 16 validators per block can partially withdraw their staked ETH. Full withdrawals, however, involve a multi-step process comprising an \emph{exit queue} and a \emph{withdrawal delay}. In the best-case scenario, the minimum time required to clear the exit queue is 5 epochs, approximately 32 minutes \cite{2023EthereumPoS}.

\subsection{The Merge}
\label{sec:Merge}

Throughout 2022 and 2023, Ethereum transitioned from the PoW to the PoS consensus mechanism \cite{2023EthereumRoadmap}. This upgrade, referred to as Ethereum 2.0 or "The Merge," was aimed at improving the network's scalability, security, and sustainability. As part of this process, the Beacon Chain, which serves as the PoS component of Ethereum 2.0, was successfully integrated into the Ethereum mainnet. Initially deployed as a separate PoS chain, the Beacon Chain has now been unified with the primary Ethereum network.

Ethereum PoW and Ethereum PoS differ fundamentally in their operation. PoW is resource-intensive, relying on miners to expend significant computational power and energy to solve complex puzzles. In contrast, PoS is far more energy-efficient, selecting validators to propose and attest to blocks based on the amount of ETH they stake as collateral. While PoW miners compete to solve puzzles and secure rewards, Etehreum validators are guaranteed periodic rewards, provided they fulfill their roles.
Additionally, Ethereum PoS offers enhanced scalability compared to PoW, as it is not constrained by hardware or energy demands, enabling more efficient transaction processing and higher throughput.

Within the Ethereum blockchain, malicious actors with dishonest intentions constantly seek vulnerabilities. This selection presents the potential attack vectors on Ethereum's security and outlines the minimum defensive thresholds of staked ETH to guard against these threats.

\subsection{Attacks on Ethereum Consensus}

It is a common misconception that a successful attacker on the Ethereum blockchain can mint new ETH tokens or steal tokens from user accounts. These actions are inherently impossible due to the strict requirements enforced by the Ethereum protocol. Every transaction on Ethereum must satisfy specific conditions, including being signed by the sender's private key and having a sufficient balance. Transactions failing to meet these criteria are automatically rejected by the network.
Instead, attackers typically target vulnerabilities related to block reorganization (reorgs), double finality, and finality delays \cite{2023EthereumAttacks, schwarzschilling2021attacks}. These exploits aim to undermine the consensus process and the integrity of the blockchain's state, rather than directly altering account balances or token issuance.

\paragraph{Reorg} 
Reorganization, or "reorg," refers to the rearrangement of blocks, involving the addition or removal of blocks from the canonical chain. There are two primary types: ex-ante and ex-post reorgs. In an ex-ante reorg, an attacker replaces a block from the canonical chain before it has been finalized. This enables manipulations such as double-spending or transaction reordering for MEV exploitation. An ex-post reorg, on the other hand, involves removing a verified block from the canonical chain, a task that requires control of over two-thirds of the staked ETH. The most extreme form of ex-ante reorg, known as finality reversion, becomes feasible when an attacker controls more than $\frac{1}{3}$ of the total staked ETH, crossing the threshold of economic finality.

\paragraph{Finality Delay} 
Finality delay occurs when an attacker with control over at least one-third of staked ETH prevents the Ethereum network from finalizing blocks. In Ethereum PoS, each block requires attestation by two-thirds of the staked ETH to reach finality. If one-third or more of the staked ETH is maliciously attesting or failing to attest, the supermajority required for finality cannot be achieved. After four epochs of failed finalization, validators that fail to attest or act maliciously are progressively slashed until their collective stake drops below the one-third threshold, restoring the supermajority. While this attack disrupts the network, it provides limited direct benefits to the attacker unless tied to specific financial incentives for disruption.

\paragraph{Double Finality} 
Double finality occurs when two forks of the Ethereum blockchain simultaneously reach finality, leading to a permanent chain split. This scenario is theoretically possible if an attacker controlling 34\% of the staked ETH performs double-voting, where the attacker simultaneously votes for two conflicting forks. This action results in the attacker's validators being slashed with the highest penalties, making the attack costly. Resolving such a chain split would require significant off-chain coordination by the Ethereum community.

\paragraph{51\% Attack} 
Attackers controlling more than 51\% of the total staked ETH can split the Ethereum blockchain into two forks of equal size and gain control over the fork choice algorithm. While they cannot alter historical transactions, they can control future blocks by directing their majority votes toward a preferred fork. This allows them to censor specific transactions and reorder blocks to extract MEV rewards, effectively controlling the blockchain's operation.

\paragraph{66\% Attack} 
Attackers with over 66\% of the total staked ETH can exert full control over the blockchain. They can finalize their preferred fork using the dishonest supermajority, enabling them to perform ex-post reorgs to alter historical blocks and execute finality reversions to control future ones. Such an attack undermines both the integrity and security of the Ethereum network.

The influence of an attacker over new blocks increases with the amount of staked ETH they control. Table \ref{tab:AttacksThresholds} outlines the minimum thresholds of staked ETH required to execute various types of attacks on the Ethereum network. The primary defense against such attacks lies in their associated cost. Any significant portion of staked ETH used in malicious activities would ultimately be slashed, especially after the social layer of Ethereum aligns with the honest minority fork.

\begin{table}[t]
\centering
\caption{Thresholds of Staked ETH Required for Various Ethereum Attack Scenarios \cite{2023EthereumAttacks, schwarzschilling2021attacks}}
\begin{tabular}{|c|l|}
\hline
\textbf{Staked ETH} & \textbf{Attack Type} \\ \hline
33\% & Delay finality \\ 
34\% & Cause double finality \\ 
51\% & Censorship and control over blockchain's future \\ 
66\% & Censorship and control over blockchain's future and past \\ \hline
\end{tabular}
\label{tab:AttacksThresholds}
\vspace{-1.5em}
\end{table}

%% file: sections/03LiquidStaking.tex
\section{Liquid Staking}
    \label{sec:liquidstaking}

Liquid Staking Tokens (LSTs) are tokenized representations of staked tokens \cite{Scharnowski2022LiquidDiscovery}. Liquid Staking Protocols (LSPs) are smart contracts that facilitate the minting and burning of LSTs. To function effectively, LSPs must manage validators, distribute staking rewards, and handle staking exit queues, among other operational challenges.

Each LST has two distinct prices: the \textit{market price}, at which the LST is traded on centralized or decentralized exchanges, and the \textit{protocol value} (also referred to as the fair or reserve value), which is the price at which the LSP mints or burns new LSTs. Arbitrageurs play a crucial role in aligning these two prices, ensuring price stability and market efficiency. 

This section introduces the architecture of LSPs, presents a related taxonomy, explores the token models of LSTs, and explains in detail how arbitrageurs maintain the peg between market price and protocol value. Finally, it summarizes the application of LSTs in DeFi and analyzes their compatibility with DeFi protocols. 

\subsection{Architecture}

\begin{figure}[tbp]
\centerline{\includegraphics[width=0.5\textwidth]{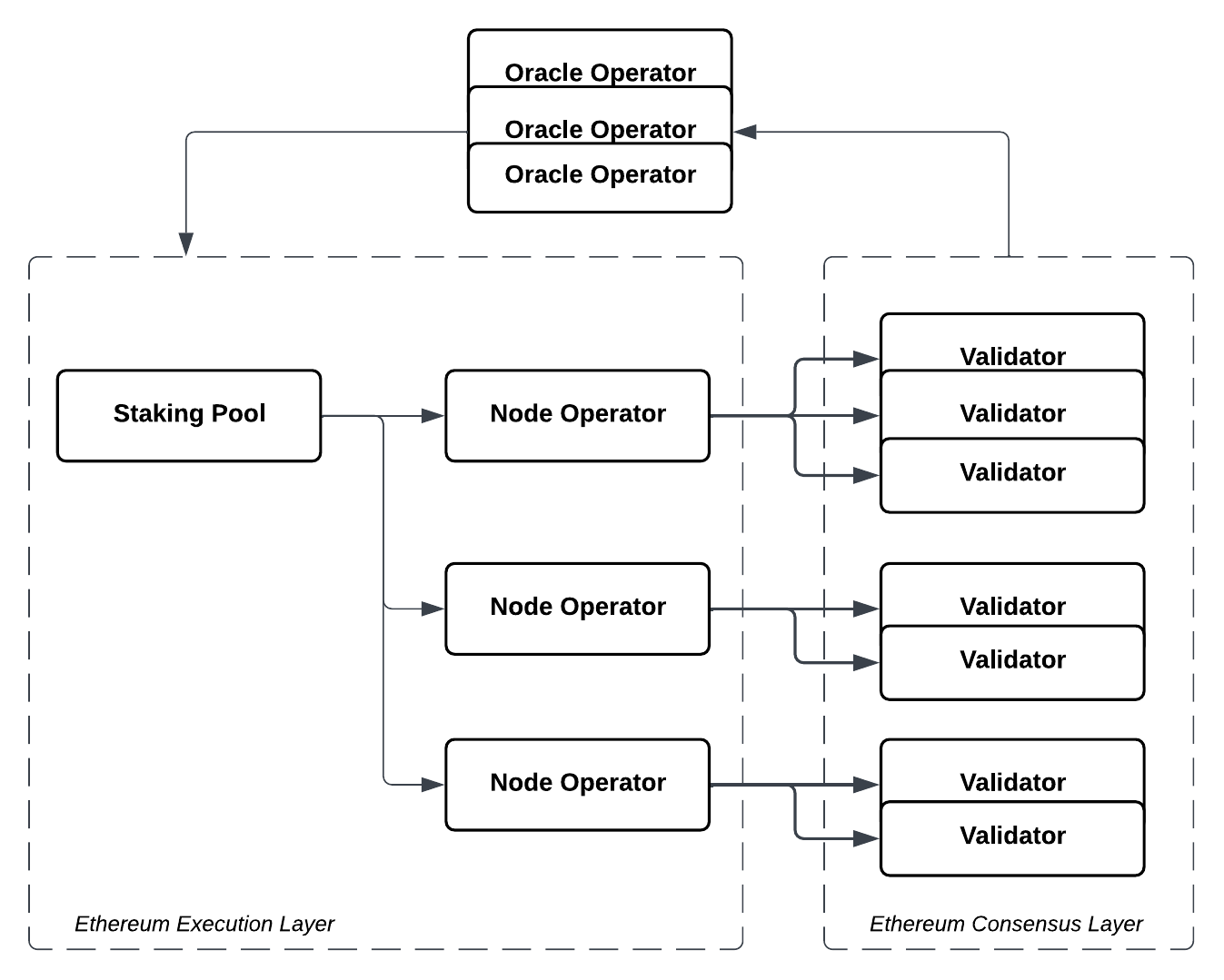}}
\caption{Overview of the Liquid Staking Protocol (LSP) architecture, including its key components: validators, node operators, staking pools, and oracle operators.}
\label{fig:LSPArchitecture}
\end{figure}

The high-level architecture of Liquid Staking Protocols (LSPs) is illustrated in Figure \ref{fig:LSPArchitecture}. It consists of validators, node operators, and the staking pool. The flow of ETH, staking reward distributions, slashing penalties, and exit queues are all managed automatically by the smart contracts of the LSP. Additionally, LSPs operating with staked ETH rely on a network of oracles to synchronize the state between Ethereum's Beacon Layer and Execution Layer.

\paragraph*{Validators} Validators are responsible for maintaining the consistency and security of the PoS blockchain. Their duties include monitoring incoming transactions, attesting to new block proposals, and verifying the legitimacy of transactions within these blocks. Periodically, validators propose new blocks. Unlike PoW miners, who compete for block rewards, validators are guaranteed rewards if they fulfill their duties. However, failure to meet obligations, such as missing an attestation or block proposal, results in penalties known as \textit{slashing}.

\paragraph*{The Staking Pool} The staking pool handles user deposits and withdrawals, as well as the distribution of staking rewards. Users mint LSTs by depositing ETH into the staking pool, while burning LSTs allows users to withdraw ETH from the pool. The LSP allocates staked ETH to node operators in a randomized manner, ensuring fairness.

\paragraph*{Node Operators} Node operators manage the infrastructure required to run validators. This includes maintaining hardware, performing software updates, monitoring performance and security, and ensuring uptime. In return for their contributions, node operators receive staking rewards. These rewards are periodically transferred from validators to the staking pool in a process known as \textit{skimming}.

\paragraph*{Oracle Operators} Oracle operators responsible for synchronizing the state between Ethereum's Beacon Layer and Execution Layer. Since these two layers lack native communication, LSPs rely on a network of oracles to ensure consistent and regular state synchronization.

\subsection{Taxonomy by Node Operator Selection}  
LSPs adopt different approaches for approving new node operators, mainly using a permissioned (whitelisting by DAO) or permissionless (with collateral) models.

\paragraph{Whitelisting} In the \textit{permissioned model}, the protocol only includes trusted node operators in its network. For instance, Lido \cite{2020Lido:Whitepaper} employs a DAO governance mechanism, where the addition of a new node operator is determined through a DAO vote.

\paragraph{Whitelisting} In contrast, the \textit{collateral-based approach} allows for a permissionless network of node operators. Here, operators must post collateral to ensure the proper performance of their validators. This model is utilized in protocols like Rocket Pool \cite{2023RocketPool:Documentation} - 8ETH, and Stader \cite{2023Stader:Documentation} - 4 ETH.
Furthermore, some LSP make the list of their validators visible, ensuring transparency, while others do not disclose this information, prioritizing operational privacy.

\begin{figure*}[ht]
\centerline{\includegraphics[width=0.7\textwidth]{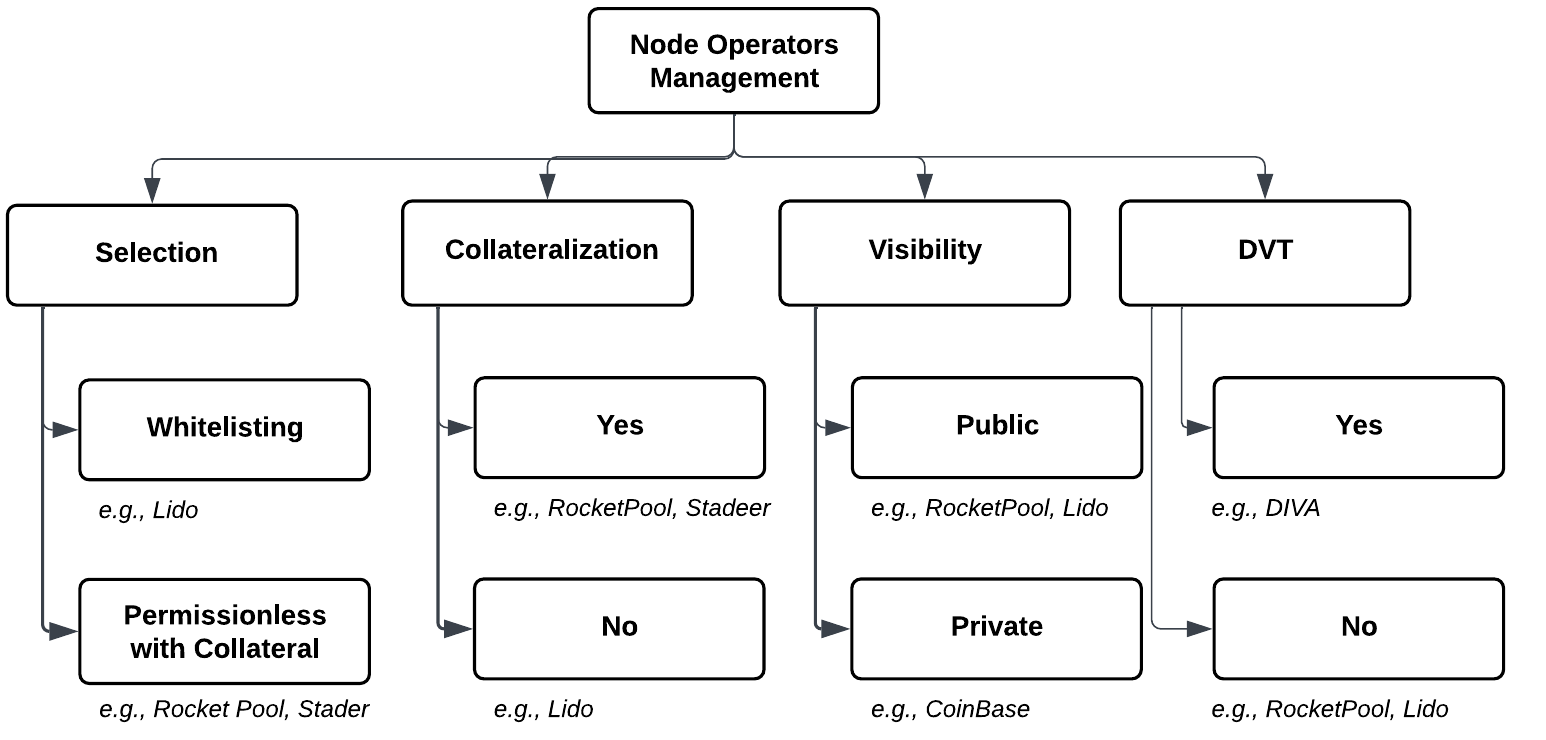}}
\caption{Taxonomy of Liquid Staking Protocols (LSPs) bases on the collaboration models with validators, showcasing differences in node operator approval mechanisms, validators visibility or applications of Distributed Validator Technology (DVT).}
\label{fig:taxonomyLSP}
\end{figure*}

\paragraph*{Distributed Validator Technology}  
Distributed Validator Technology (DVT) \cite{2023EthereumDVT} is a novel approach designed to enhance the security and resilience of validators by distributing key management and signing responsibilities among multiple node operators. Its primary objective is to mitigate risks associated with a single point of failure, where one node operator holds the private keys for a validator. In the DVT model, the complete private key is never concentrated on a single machine; instead, it is distributed across multiple node operators organized into a cluster. This ensures that the validator can continue operating even if certain nodes within the cluster go offline (\emph{liveness}) or if some node operators act maliciously (\emph{fault tolerance}).

\begin{figure*}[!tbp]
\centerline{\includegraphics[width=\textwidth]{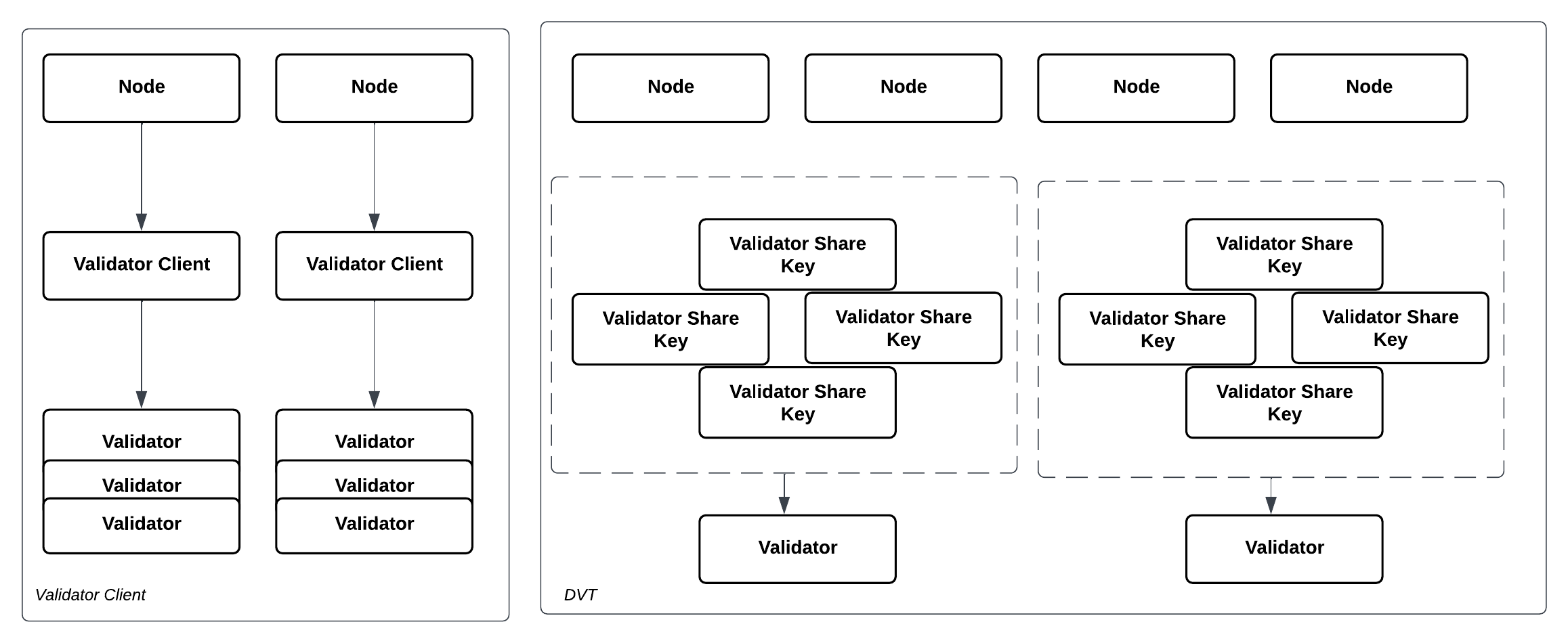}}
\caption{Comparison of Validator Key Management with and without Distributed Validator Technology, highlighting enhanced security and resilience.}
\label{fig:DVTArchitecture}
\end{figure*}

Validators generate two sets of public-private key pairs: one for participating in consensus protocols and another for accessing withdrawal funds. While withdrawal keys can be securely stored in cold storage, validator private keys must remain online continuously. A compromise of the validator private key could enable an attacker to control the validator, leading to slashing penalties or the loss of staked ETH. DVT provides a robust solution to this vulnerability.

With DVT, LSPs' node operators can participate in staking while ensuring the validator's private key remains securely stored in cold storage. This is achieved by encrypting the original complete validator key and dividing it into key shares. These key shares are stored online and distributed across multiple nodes, enabling the distributed operation of the validator. The full, original master validator key remains securely stored offline.

When validator management is distributed across multiple operators and machines, it becomes resilient to individual hardware or software failures, significantly reducing downtime risk. Additional resilience can be achieved by employing diverse hardware and software configurations across the nodes within a cluster. In the event of a node operator failure within a cluster, the remaining nodes ensure the validator continues operating seamlessly.

LSPs define policies for distributing key shares among node operators, considering factors such as random allocation or prioritization of smaller nodes. The potential risks and implications of these policies are analyzed in Appendix \ref{sec:ETHSecurity}.

\paragraph{Impact on Ethereum Security}  
In the permissionless model of node operators, anyone can set up validator for LSPs. Annonymous node operators are only required to deposit collateral, with the remaining ETH supplied by the staking pool of the LSP. This reduces the amount of ETH attackers need to operate a validator, as they only need to cover the collateral requirements set by the LSP. Consequently, this leads to a decreased attack threshold for compromising the Ethereum network, as shown in Table \ref{tab:LSDAttackThreshold}.

Further reductions in the attack threshold are possible with naively implemented DVT, particularly without random key distribution. If only 2/3 of the key shares are required to operate a validator and node operators can select their key shares, the attack threshold decreases further by a factor of 2/3. However, LSPs can mitigate this risk by implementing random distribution of key shares to validator private keys among node operators.

\begin{table}[t]
\centering
\caption{Ethereum minimum security attack thresholds with varying collateral to operate a validator, without and with DVT.}
\begin{tabular}{|r|r|r|}
\hline
\textbf{Collateral} & \textbf{Threshold (\%)} & \textbf{Threshold with DVT (\%)} \\ 
\hline
32 ETH & 33.00 & 22.00 \\ 
16 ETH & 16.50 & 11.00 \\ 
8  ETH & 11.00 & 7.30 \\ 
4  ETH & 4.13  & 2.75 \\ 
1  ETH & 1.03  & 0.69 \\ 
\hline
\end{tabular}
\label{tab:LSDAttackThreshold}
\end{table}

\subsection{Taxonomy by Token Models}  
One of the most significant aspects of Liquid Staking Tokens (LSTs) is their token model, which determines how staking rewards (and slashing penalties) are distributed among token holders. There are two major types of LST token models: \emph{rebase} and \emph{reward-bearing}. A third model, the \emph{dual token model}, is becoming increasingly obsolete.

\begin{figure}[ht]
\centerline{\includegraphics[width=0.35\textwidth]{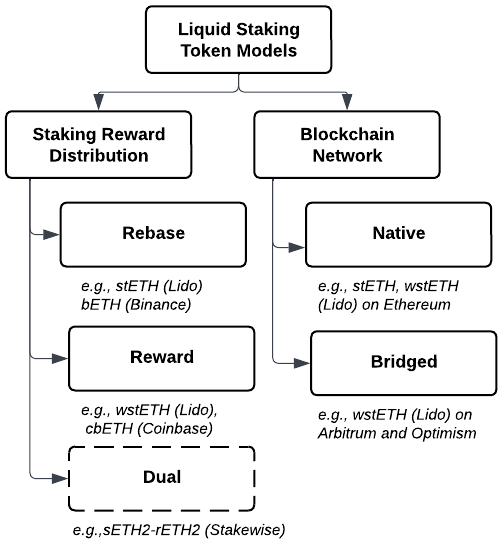}}
\caption{Taxonomy of Liquid Staking Tokens categorized by staking reward redistribution models and blockchain network deployment. The dual-token model is becoming obsolete.}
\label{fig:taxonomyLST}
\end{figure}

\paragraph{Rebase Tokens}  
Rebase-LSTs maintain a 1:1 peg to ETH, with the protocol periodically increasing the token supply to reflect the staking rewards earned. An example of a rebase LST is \textit{stETH}, minted by Lido. Lido regularly increases the balance of stETH held by users to mirror the additional ETH earned as staking rewards. The primary advantage of rebase LSTs is their straightforward design. However, they face notable limitations, including restricted compatibility with most DeFi protocols—especially decentralized exchanges (DEXs) and lending platforms—and the inability to bridge to other blockchains. These limitations are discussed further in subsequent subsections.

\paragraph{Reward-Bearing Tokens}  
Reward-LSTs continuously increase in value to reflect the accumulation of staking rewards, rather than maintaining a 1:1 peg to ETH. This design allows their value to grow relative to ETH over time. Reward-LSTs are fully compatible with DeFi and can be bridged across blockchains. Many projects adopt this model, including RocketPool \cite{2023RocketPool:Documentation}, Stader \cite{2023Stader:Documentation}, and CoinBase Wrapped Staking \cite{2022CoinBaseETH:Whitepaper}. Some LSPs, like Lido, DIVA \cite{2023DIVA:Documentation}, and Binance \cite{2023BinanceWBETH:Documentation}, offer both rebase and reward-bearing tokens.

\paragraph{Dual Token Model}  
The dual token model separates the asset token from the staking rewards it generates. For example, StakeWise in its initial desing \cite{2023StakeWise:Documentation} issued two tokens: \textit{sETH2}, which maintains a 1:1 peg to ETH, and \textit{rETH2}, which accumulates value based on staking rewards. The primary drawback of this model is the fragmentation of liquidity between the two separate tokens, which complicates usability and integration within DeFi ecosystems. Consequntly, this token model is disappearing.

\paragraph{Native vs. Bridged Tokens}  
Layer-2 blockchains (L2s) are becoming increasingly popular due to their ability to leverage Ethereum's security while offering significantly lower gas fees compared to the Ethereum mainnet \cite{gangwal2022survey, yee2022shades}. LSTs available on L2s are often bridged tokens originating from the Ethereum mainnet. In this model, the bridge locks LST tokens on the Ethereum mainnet and mints equivalent wrapped tokens on the L2 chain. While effective, this approach is vulnerable to bridge hacking attacks and is limited to supporting only reward-based LSTs \cite{2023LidoRebase}. Moreover, only reward-based LSTs are capable of being bridged.

An alternative approach employs native tokens on L2 networks. Unlike bridged tokens, native LSTs directly represent tokenized staked assets on the Ethereum mainnet without requiring wrapping. Achieving native LSTs on L2s involves implementing proof-of-storage protocols that provide L2 networks with reliable information about the state of assets on the Ethereum mainnet. This ensures a seamless and efficient connection between the two layers, improving security and usability \cite{2023ChainLinkCrossChainLSD}.


\subsection{LSTs as DeFi Pegged Tokens}
\label{sec:Peg}

Each LST is associated with a \emph{protocol value}, also known as fair value, peg value, or reserve value. This value is calculated by dividing the total reserves, including staked ETH and protocol reserves, by the number of tokens in circulation. The \emph{market value} of an LST is the price at which it is traded on DEXs or CEXs. Discrepancies between the protocol value and market value may arise due to market inefficiencies, creating arbitrage opportunities.

There are two primary methods to acquire LSTs:
\begin{itemize}
    \item Purchase LSTs on a DEX or CEX at the market price.
    \item Mint LSTs directly from the LSP at the fair value.
\end{itemize}

The processes for minting and burning LSTs are detailed in Algorithms \ref{alg:minting} and \ref{alg:burning}, respectively. Users who mint LSTs, referred to as Depositors or Stakers, deposit ETH into the staking pool of the LSP in exchange for LSTs. The value of the newly minted LSTs matches the value of the deposited ETH, maintaining the protocol value. Once enough ETH is accumulated in the staking pool to create a new validator, the LSP distributes ETH to randomly selected node operators. These node operators create a new validator and stake the ETH. 

In permissionless LSPs, node operators must provide collateral (e.g., 4 ETH or 8 ETH) to create a new validator. The remaining ETH required to meet the 32 ETH threshold is supplied by the staking pool. If no node operators are available to provide the collateral, LSTs for Depositors may be acquired on DEXs. 

The timing of staking rewards depends on the LSP implementation. For example, in Lido, staking rewards are collected within 24 hours of depositing ETH, regardless of whether the ETH has been staked. This approach spreads the risk of staking queues among all LST holders.

\begin{algorithm}
\caption{Minting LST by LSP}
\label{alg:minting}
\begin{algorithmic}[1]
\STATE \textbf{Collect} ETH from the Depositor into the staking pool.
\STATE \textbf{Mint} new LSTs equivalent to the value of the deposited ETH and provide them to the Depositor.
\STATE When sufficient ETH accumulates in the staking pool, distribute it to a randomly selected node operator to \textbf{stake} by creating a new validator.
\end{algorithmic}
\end{algorithm}

Similarly, there are two methods to redeem LSTs:
\begin{itemize}
    \item Sell LSTs on a DEX or CEX at the market price.
    \item Burn LSTs directly within the LSP for a fair price.
\end{itemize}

If sufficient ETH is available in the staking pool, there is no need to unstake ETH, as described in Algorithm \ref{alg:burning}. After burning LSTs, the equivalent value of ETH is paid to the Depositor. If the staking pool lacks sufficient ETH, the LSP requests node operators to unstake ETH. The unstaked ETH is then allocated to the staking pool to pay the Depositor. Depending on the implementation, LSPs may or may not have mechanisms to force node operators to unstake ETH. Exit queues for unstaking ETH could further delay the redemption process for Depositors.

\begin{algorithm}
\caption{Burning LST by LSP}
\label{alg:burning}
\begin{algorithmic}[1]
\STATE \textbf{Collect} LSTs from the Depositor and exchange them for an equivalent value of ETH from the staking pool, if available.
\STATE \textbf{Burn} the collected LSTs.
\STATE If the staking pool lacks sufficient ETH, request node operators to \textbf{unstake} ETH equivalent to the value of the collected LSTs.
\end{algorithmic}
\end{algorithm}

\paragraph*{Arbitrage Opportunities}

Market inefficiencies can lead to disparities between the protocol value and market value of LSTs. Arbitrage strategies aim to exploit these differences to equalize the protocol and market values, as outlined in Algorithms \ref{alg:arbitrageLow} and \ref{alg:arbitrageHigh}.

When the protocol value of LSTs exceeds the market price, arbitrageurs can profit by purchasing LSTs on a DEX at the lower market price and then burning them within the LSP to redeem the higher protocol value.

\begin{algorithm}
\caption{Arbitrage Strategy for Market Price $<$ Protocol Price}
\label{alg:arbitrageLow}
\begin{algorithmic}[1]
\STATE \textbf{Buy} undervalued LSTs on a DEX at the market price.
\STATE \textbf{Burn} the LSTs at the LSP to redeem the protocol price.
\end{algorithmic}
\end{algorithm}

Conversely, when the market price of LSTs exceeds the protocol value, arbitrageurs mint LSTs within the LSP at the protocol price and sell them on a DEX at the higher market price.

\begin{algorithm}
\caption{Arbitrage Strategy for Market Price $>$ Protocol Price}
\label{alg:arbitrageHigh}
\begin{algorithmic}[1]
\STATE \textbf{Mint} LSTs at the LSP for the protocol price.
\STATE \textbf{Sell} the minted LSTs on a DEX at the market price.
\end{algorithmic}
\end{algorithm}

It must be noted, however, that not every price disparity creates an arbitrage opportunity. Some discrepancies persist due to scalability limitations of LSPs. For example, LSPs with permissioned node operators may face challenges in attracting sufficient operators to handle new incoming tokens. Factors such as transaction costs on DEXs, gas fees, and staking or unstaking queues contribute to these discrepancies persisting over time, as empirically studied in later sections of this work.

\subsection{Liquid Staking Tokens in DeFi}
Decentralized Finance (DeFi), blockchain-based financial services~\cite{Auer2023technologydefi,Gogol2023SoK:Risks,Werner2021SoK:DeFi}, offers an extensive array of services, including trading, lending, and digital asset management~\cite{Schar2021DecentralizedMarkets}.
The advantage of LSTs over native staking is the fact that they can be freely transferred and used in various DeFi protocols. Reward-based LSTs are more popular in DeFi because rebase tokens are not widely compatible with most protocols, as explained below. This section further presents the main uses of reward-based LSTs in DeFi.

\paragraph{Trading}  
Automated Market Markers (AMMs) that are the backbone of DEX allow and trade, list or provide liquidity to any token pairs~\cite{Xu2021SoK:Protocols}. 
LSTs can be traded on DEXs and CEXs at market prices at any time without exit queues that are an inherent component of native staking. However, the market price of LST on the exchange may differ from the protocol prices. This price dispcrepancies might lead to the arbitrage opportunities, presented earlier in this chapter. 
Most AMM-based DEXs, including the largest DEX - Uniswap, do not support trading of rebase LSTs~\cite{2023LidoRebase}. These tokens can be traded on Curve DEX~\cite{2023LidoRebase}. Table \ref{tab:LiquidStakingInAMM} depicts how LSTs should be listed at AMM-DEX in order to minimize trading fees and optimize capital efficiency~\cite{gogol2024LSTinAMM}.

\begin{table}[t]
\centering
\caption{Optimal AMMs for Trading Pairs with Various LST Types Based on Capital Efficiency. Concentrated Liquidity Market Maker (CLMM)~\cite{Adams2021UniswapCore} Requires Periodic Rebalancing~\cite{gogol2024LSTinAMM}.}
\begin{tabular}{|l|l|l|}
\hline
\textbf{Token 1} & \textbf{Token 2} & \textbf{AMMs} \\ \hline
rebase-LST       & ETH              & Stableswap~\cite{Egorov2019StableSwap-efficientLiquidity}    \\ 
rebase-LST       & rebase-LST       & Stableswap    \\ 
reward-LST       & ETH              & Cryptoswap~\cite{Egorov2021AutomaticPeg}, CLMM(*) \\ 
reward-LST       & reward-LST       & Cryptoswap, CLMM(*)    \\ \hline
\end{tabular}
\label{tab:LiquidStakingInAMM}
\vspace{-1.5em}
\end{table}

\paragraph{Collateral in Lending}  
LSTs can be used as collateral in DeFi lending protocols to borrow other tokens~\cite{gudgeon2020defiloan}. Reward-based LSTs are particularly favored as their value appreciates over time due to staking rewards, making them a dominant collateral asset in DeFi, surpassing ETH~\cite{2023MessariQ123Report}. Rebase-LSTs, however, are not used as collateral because staking rewards would need to be divided between the borrower and the lender. In contrast, for reward-LSTs, all staking rewards accrue to the collateral provider.

\paragraph{Leveraging}  

\begin{algorithm}
\caption{Leveraging Strategy for LSTs}
\label{alg:LSTLeverage}
\begin{algorithmic}[1]
\STATE \textbf{Deposit} LSTs as collateral in a lending protocol.
\STATE \textbf{Borrow} ETH against the collateralized LSTs.
\STATE \textbf{Buy} additional LSTs using the borrowed ETH on a DEX.
\STATE \textbf{Repeat} steps 1-3 to increase leverage, ensuring the collateralization ratio remains above the protocol's liquidation threshold.
\end{algorithmic}
\end{algorithm}

LSTs can be leveraged in lending protocols~\cite{mueller2024defi,wang2022speculative,heimbach2023defileverage} through a process known as \emph{looping}~\cite{2024ETHSaver}. This strategy, depicted in Algorithm~\ref{alg:LSTLeverage}, involves using LSTs as collateral to borrow ETH, which is then used to purchase additional LSTs. 

\begin{algorithm}
\caption{Leveraging LSTs Strategy with Flash Loan}
\label{alg:LSTFlashLoan}
\begin{algorithmic}[1]
\STATE \textbf{Borrow} a flash loan in ETH from a lending protocol.
\STATE \textbf{Buy} LSTs with the borrowed ETH on a DEX.
\STATE \textbf{Deposit} the purchased LSTs as collateral in a lending protocol.
\STATE \textbf{Borrow} ETH against the collateralized LSTs.
\STATE \textbf{Repay} the flash loan using the borrowed ETH.
\end{algorithmic}
\end{algorithm}

The more capital-efficient leveraging method applies flash loans. In this strategy, presented in Algorithm~\ref{alg:LSTFlashLoan} a flash loan is used to purchase LSTs, lock them as collateral in a lending protocol, borrow ETH, and repay the loan. According to emirical research, most LST leveraging strategies gained higher returns compared to native staking~\cite{xiong2024ExpoLSD}. 
However, the DeFi leverage is exposed to liquidation risks~\cite{mueller2024defi,wang2022speculative,heimbach2023defileverage}.

\paragraph{Providing Liquidity to AMM-DEX}  
LSTs can be utilized in liquidity provision to AMM-based DEXs~\cite{Fritsch2021ConcentratedMakers,Heimbach2022RisksProviders}. In LST-ETH liquidity pools, part of the LSTs is converted to ETH, meaning only a fraction of the liquidity pool earns staking rewards. High trading volumes and corresponding swap fees are necessary to compensate for missed staking opportunities. Since future LST prices can be estimated based on the current staking rate, LSTs are frequently traded in AMMs with concentrated liquidity models, such as Uniswap v3. However, provided liquidity to AMM pool with LSTs does not always provide higher rewards than native staking due to impermanent loss~\cite{Fritsch2021ConcentratedMakers,Heimbach2022RisksProviders}. The topic of profitability of various DeFi strategies with LSTs is further discussed in Section~\ref{sec:relatedwork}, as empirical research showed that most of LP-strategies involving LSTs leads to the lower returns than native staking~\cite{gogol2024LSTinAMM,xiong2024ExpoLSD}. 

\paragraph{Collateral for Stablecoins}  
Decentralized stablecoins are DeFi tokens that maintain their value pegged to a fiat currency or commodity thanks to the on-chain assets that backed them~\cite{2019stablecoins,ANTE2021101867}. 
LSTs, as well as LP-positions in AMMs including LSTs, can be provided as collateral to mint decentralized stablecoins\cite{2024Ethena:Documentation,2024Lido:DAI}. The appreciating value of LSTs, driven by staking rewards, reduces the liquidation risk of the collateral over time, enhancing stability and efficiency in stablecoin systems.

\paragraph{Operating a Validator with Less than 32 ETH}  
Operating an Ethereum validator typically requires a deposit of 32 ETH. However, with LSPs, it is possible to operate a validator with only 4 ETH~\cite{2023Stader:Whitepaper} or 8 ETH~\cite{2023RocketPool:Documentation}, with the remaining ETH provided by the staking pool of the LSP.

The risks inherent in LST-based DeFi, particularly in scenarios where LST de-pegs result in liquidations~\cite{qin2021empirical}, are becoming increasingly prominent in the literature and are discussed in Section~\ref{sec:relatedwork}.

%% file: sections/05LST_Protocols.tex
\section{Comparison of Liquid Staking Protocols}
    \label{sec:empirical}

Liquid staking is the largest category among DeFi protocols by TVL and the preferred method of staking for users. Currently, 37\% of all staked ETH is held via Liquid Staking Tokens (LSTs). Table \ref{tab:LiquidStakingProtocolTVL} lists the major Liquid Staking Protocols (LSPs), their total value locked (TVL), market share within the broader ETH staking market, and other relevant metrics. 

\begin{table*}[t]
\centering
\caption{Overview of major Liquid Staking Protocols (LSPs), highlighting total value locked (TVL), market share of total staked ETH, governance (Gov.) token and associated metrics.}
\label{tab:LiquidStakingProtocolTVL}
\begin{tabular}{|l|l|r|r|l|p{1.7cm}|r|r|r|}
\hline
\textbf{Protocol} &  & \textbf{TVL (\$B)} & \textbf{Market (\%)} & \textbf{LSTs} & \textbf{Gov. Token} & \textbf{Nodes} & \textbf{Validators} & \textbf{Fees (\%)} \\ 
\hline
Lido                  & \cite{2020Lido:Whitepaper, 2023Lido:Documentation} & 32.49 & 28.00 & stETH, wstETH & LDO  & 29   & 276k  & 10   \\ 
Rocket Pool           & \cite{2023RocketPool:Documentation}               & 2.48  & 2.19  & rETH          & RPL  & 3088 & 25k   & 20   \\ 
Binance Staked ETH    & \cite{2023BinanceETH:Documentation}               & 5.91  & 5.16  & bETH, wbETH   & -    & -    & -     & 10   \\ 
Frax Ether            & \cite{2023FraxETH:Documentation}                 & 0.42  & 0.38  & sfrxETH       & -    & -    & 7k    & 10   \\ 
CoinBase Wrapped ETH  & \cite{2022CoinBaseETH:Whitepaper}                & 0.59  & 0.51  & cbETH         & -    & -    & -     & 25   \\ 
StakeWise             & \cite{2023StakeWise:Documentation}               & 0.49  & 0.39  & sETH2, rETH2  & SWISE & 4    & 3k    & 10   \\ 
Stader                & \cite{2023Stader:Documentation}                  & 0.69  & 0.60  & ETHx          & SD   & 109  & 718   & 10   \\ 
Swell                 & \cite{2023Swell:Documentation}                   & 0.24  & 0.18  & swETH         & SWELL & 8    & -     & 10   \\ 
Liquid Collective     & \cite{2023LiquidCollective:Documentation}        & 0.35  & 0.28  & lcETH         & -    & -    & -     & 10   \\ 
Ankr                  & \cite{2023Ankr:Documentation}                    & 0.05  & 0.01  & ankrETH       & ANKR & -    & -     & 10   \\ 
DIVA                  & \cite{2023DIVA:Documentation}                    & 0.03  & 0.01  & divETH, wdivETH & DIVA & -    & -     & 10   \\ 
\hline
\end{tabular}
\end{table*}

Lido is the dominant LSP, accounting for over 31\% of staked ETH, with a TVL of \$14.44 billion. It operates through 29 node operators who collectively manage approximately 276,000 validators—an average of 9,500 validators (1.07\% of total staked ETH) per Lido node operator. In contrast, RocketPool, with 3,088 node operators, demonstrates a significantly more decentralized structure, with each operator managing an average of eight validators.

In the subsequent sections, we analyze the design decisions of various LSPs and empirically study the stability of their peg to ETH staking rewards. This section concludes with a comparative analysis of the historical performance of LSTs against native staking.


\subsection{Major Liquid Staking Protocols}  
This section discusses the architecture of major Liquid Staking Protocols (LSPs). The selected protocols were chosen based on the value of staked ETH \cite{2022DeFiLlama} and represent all previously described mechanisms. A summary of the design decisions for these LSPs is presented in Table \ref{tab:LiquidStakingProtocols}. 

\begin{table}[t]
\centering
\caption{Taxonomy of Liquid Staking Protocols (LSPs) comparing permissionless (open) access to operate a validator, collateral requirements to run operate a validator and visibility of validator sets.}
\label{tab:LiquidStakingProtocols}
\begin{tabular}{|l|c|l|c|}
\hline
\textbf{Protocol} & \textbf{Permissionless} & \textbf{Collateral} & \textbf{Visibility} \\ 
\hline
Lido                  & \ding{55} & -           & \ding{51} \\ 
Rocket Pool           & \ding{51} & 8 ETH       & \ding{51} \\ 
Coinbase Wrapped ETH  & \ding{55} & -           & \ding{55} \\ 
Binance Staked ETH    & \ding{55} & -           & \ding{55} \\ 
Frax Ether            & \ding{55} & -           & \ding{51} \\ 
Ankr                  & \ding{55} & -           & \ding{51} \\ 
StakeWise             & \ding{55} & -           & \ding{51} \\ 
Stader                & \ding{51} & 4 ETH       & \ding{51} \\ 
Swell                 & \ding{55} & -           & \ding{51} \\ 
Liquid Collective     & \ding{55} & -           & \ding{51} \\ 
DIVA                  & \ding{51} & 1 ETH       & \ding{51} \\ 
\hline
\end{tabular}
\end{table}

Only RocketPool, Stader, and DIVA maintain a permissionless network of node operators, requiring collateral from each operator to manage validators. Consequently, RocketPool and Stader have the highest number of node operators, with over 3,000 and 100, respectively, followed by Lido with 29 nodes. Lido approves its node operators via a DAO vote.  
Another key differentiation between LSPs lies in private key management. Only DIVA supports the Distributed Validator Technology (DVT) model, where each validator's private key is divided among multiple node operators, ensuring no single operator has complete control over a validator. In contrast, RocketPool, Stader, Lido, and other LSPs give each node full control over their respective validators.

\paragraph*{Lido}  
Launched in December 2020, Lido was the first LSP. It operates as a decentralized autonomous organization (DAO) governed by Lido governance token holders. Key protocol decisions, including fee structures, staking parameters, and upgrades, are made through community voting. Lido selects its node operators based on their track record and reputation via a DAO vote. The protocol provides real-time monitoring of validator performance and staking metrics. Lido issues two LSTs for staked ETH: \textit{stETH}, which distributes staking rewards via the rebase model, and \textit{wstETH}, which operates as a reward-based token. According to CoinGecko, \textit{stETH} is the 8th largest cryptocurrency by market capitalization. Lido's market share of 31\% in staked ETH raises concerns about its impact on Ethereum blockchain security, discussed in the next section.

\paragraph*{RocketPool}  
Founded in 2016 and launched in 2021, RocketPool provides \textit{rETH}, a reward-based token. The protocol is permissionless, allowing any node operator to join the network by providing the required collateral. After the Atlas upgrade in 2023, node operators can choose between 8 ETH or 16 ETH as collateral for each validator, with the remaining 24 ETH or 16 ETH, respectively, supplied by RocketPool's staking pool. Additionally, the protocol requires a 10\% deposit of governance tokens (RPL) proportional to the ETH provided by the staking pool.

\paragraph*{Stader}  
Stader is another example of an LSP with a permissionless setup for node operators. The required collateral for node operators is just 4 ETH, significantly lower than RocketPool's 8 ETH.

\paragraph*{Stakewise}  
Stakewise operates a dual-token model: \textit{sETH2}, pegged to ETH, and \textit{rETH2}, which accumulates staking rewards. However, Stakewise has announced plans to phase out the dual model in favor of a reward-based LST.

\paragraph*{Frax}  
Frax offers a reward-based LST called \textit{sfrxETH}, which is minted after depositing \textit{frxETH}, an ETH wrapper with a 1:1 value to ETH. Unlike rebase-based LSTs, \textit{frxETH} is an ERC20 representation of ETH.

\paragraph*{Liquid Collective}  
Liquid Collective focuses on institutional customers by providing transparency around node operators. Mandatory anti-money laundering (AML) checks for users and operators ensure regulatory compliance.

\paragraph*{DIVA}  
The DIVA protocol utilizes Distributed Validator Technology (DVT) to operate Ethereum validators. Node operators use DVT Key Shares instead of validator private keys. Each validator is managed by 16 DVT Key Shares, with node operators locking one \textit{divETH} collateral per key share. The remaining 16 ETH required to run a validator is provided by DIVA's staking pool. Key shares are generated using Multi-Party Computation (MPC) technology, ensuring that private keys are never reconstructed, eliminating single points of failure.

\paragraph*{Binance and Coinbase}  
These leading centralized exchanges (CEXs) offer native staking services to their customers and have launched their own LSTs. Unlike permissionless or DAO-based LSPs, the selection of node operators is a centralized process managed internally by the exchanges and the set of node operators and validators is private. 

\subsection{Empirical Analysis}
This section examines the performance of LSTs compared to native staking of ETH. To determine daily LST performance, market price data was sourced directly from Uniswap \cite{Adams2021UniswapCore}. The analysis considers the allocation of one ETH, either to LSTs or to native staking. For staking rewards, it is assumed that rewards are claimed and re-staked daily. Staking yield rates fluctuate, as shown in Figure \ref{fig:apy}, and vary among validators. Historical staking rates were obtained from the Ethereum Explorer to ensure accuracy.

\begin{figure*}[!tbp]
  \centering
  \begin{subfigure}[t]{0.8\linewidth}
    \centering
    \includegraphics[width=\linewidth]{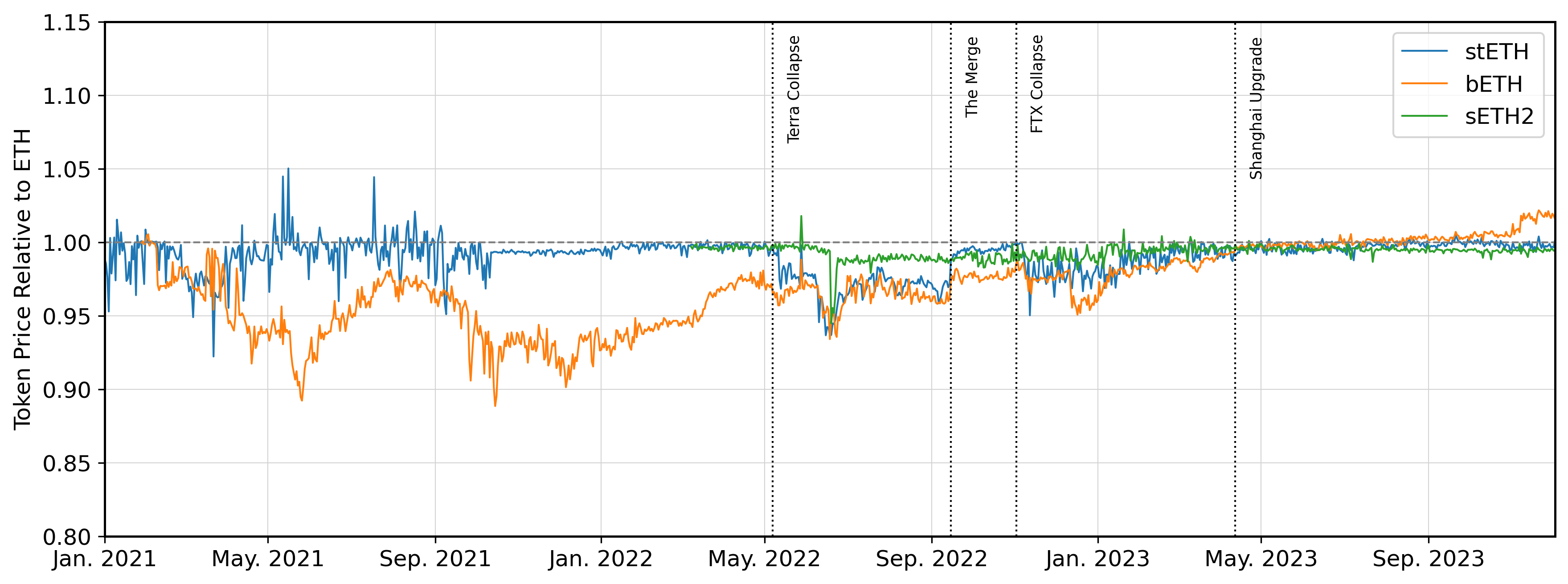}
    \caption{Market values of rebase LSTs (stETH from Lido, bETH from Binance, and sETH2 from StakeWise) during Ethereum's transition from PoW to PoS, concluded in the Shanghai Upgrade. ETH serves as the reference currency.}
    \label{fig:rebase_inception}
  \end{subfigure}

  \vspace{0.5cm} 

  \begin{subfigure}[t]{0.8\linewidth}
    \centering
    \includegraphics[width=\linewidth]{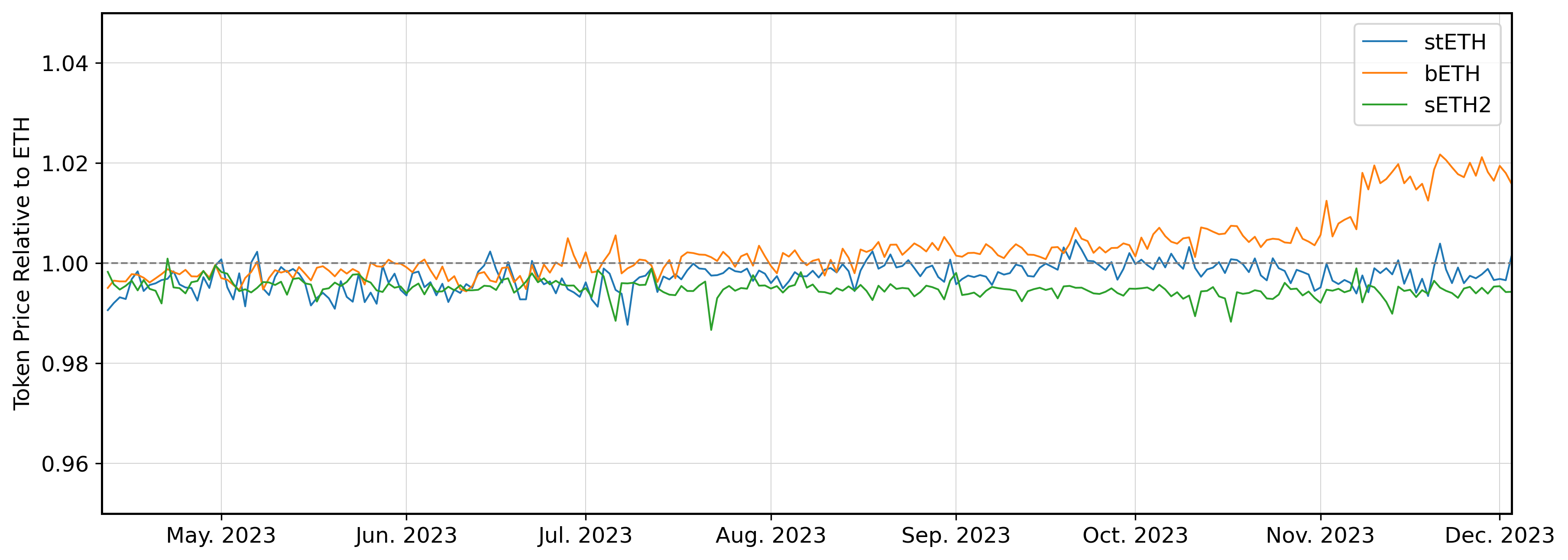}
    \caption{Market values of rebase LSTs (stETH from Lido, bETH from Binance, and sETH2 from StakeWise) after the Shanghai upgrade. ETH serves as the reference currency.}
    \label{fig:rebase_shanghai}
  \end{subfigure}

  \vspace{0.5cm} 

  \begin{subfigure}[t]{0.8\linewidth}
    \centering
    \includegraphics[width=\linewidth]{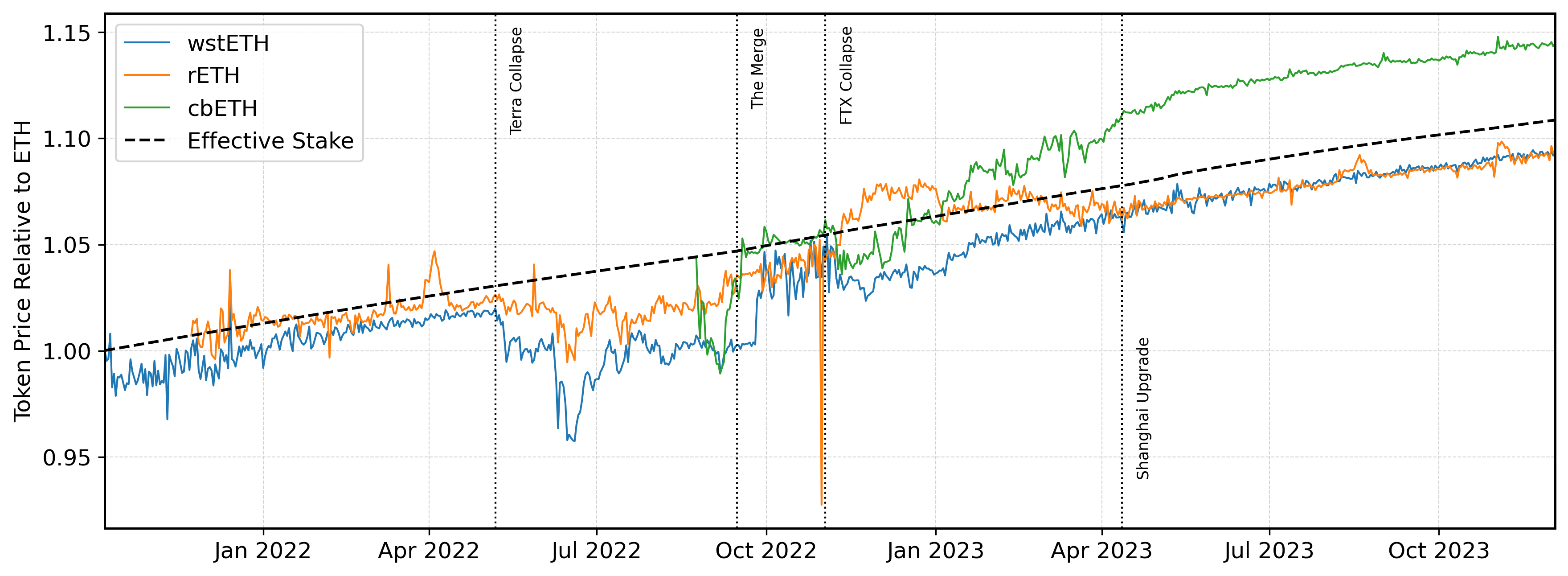}
    \caption{Market values of major reward-based LSTs during Ethereum's transition from PoW to PoS, concluded in the Shanghai Upgrade. The dotted line represents the cumulative staking rewards over time.}
    \label{fig:reward_inception}
  \end{subfigure}

  \caption{Historical performance of Liquid Staking Tokens (LSTs). (a) Rebase LSTs' market values since inception, using ETH as the reference currency. (b) Rebase LSTs' market values after the Shanghai upgrade, with ETH as the reference currency. (c) Reward-based LSTs' market values since inception, highlighting cumulative staking rewards with a dotted line.}
  \label{fig:combined_LST}
\end{figure*}

Rebase-LSTs maintain a 1:1 peg to ETH. Historically, their market values (Figure \ref{fig:rebase_inception}) have temporarily deviated from the peg, especially during periods of market turmoil. The historical performance of reward-LSTs is depicted in Figure \ref{fig:reward_inception}, which shows the compounded daily returns of reward-LSTs compared to directly staking ETH.

\paragraph*{LST Classification}
Table \ref{tab:LiquidStakingTokens} presents the taxonomy of the major LSTs that were empirically analyzed in this study, with most tokens operating in the reward-based model and naively launched on Ethereum. The differentiations between rebase and reward models for staking reward distribution  is crucial for correct empirical analysis of the tokens. 

\begin{table}[tbh]
\caption{Token taxonomy for Liquid Staking Tokens (LSTs) with staking reward distribution models supported blockchains. The analyzed tokens are naively minted only on Ethereum and bridged to other blockchain.}
\label{tab:LiquidStakingTokens}
\centering
\begin{tabular}{|l|l|l|l|}
\hline
\textbf{Protocol}     & \textbf{LST}       & \textbf{Model} & \textbf{Blockchain} \\ 
\hline
Lido                  & stETH              & Rebase         & Ethereum   \\ 
Lido                  & wstETH             & Reward         & Ethereum, Arbitrum, \\ 
                      &                    &                & Optimism, Polygon  \\ 
Rocket Pool           & rETH               & Reward         & Ethereum, Optimism, \\ 
                      &                    &                & Arbitrum            \\ 
Coinbase              & cbETH              & Reward         & Ethereum            \\ 
Binance               & bETH               & Rebase         & Ethereum             \\ 
Binance               & wbETH              & Reward         & Ethereum             \\ 
Frax                  & sfrxETH            & Reward         & Ethereum            \\ 
Ankr                  & ankrETH            & Reward         & Ethereum            \\ 
StakeWise             & sETH2       & Dual           & Ethereum           \\ 
              &  rETH2       &             &             \\ 
Stader                & ETHx               & Reward         & Ethereum           \\ 
Swell                 & swETH              & Reward         & Ethereum          \\ 
Liquid Collective     & lcETH              & Reward         & Ethereum            \\ 
DIVA                  & divETH             & Rebase         & Ethereum         \\ 
DIVA                  & wdivETH            & Reward         & Ethereum          \\ 
\hline
\end{tabular}
\end{table}

\paragraph*{Dispersion Analysis}

\begin{figure*}[htb]
  \begin{minipage}[c]{0.49\textwidth}
    \includegraphics[width=0.99\linewidth]{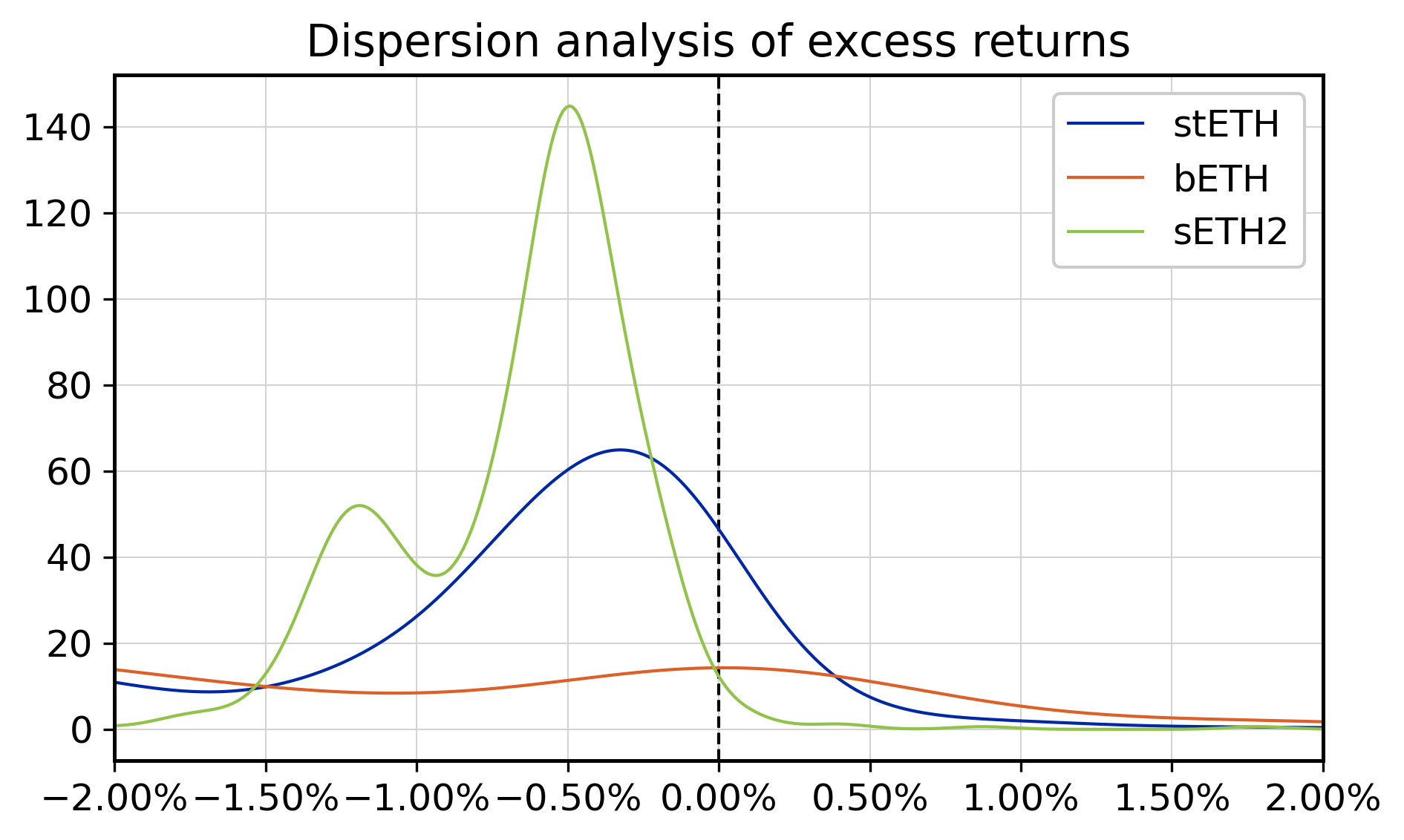} 
  \end{minipage} 
  \begin{minipage}[c]{0.49\textwidth}
    \includegraphics[width=.99\linewidth]{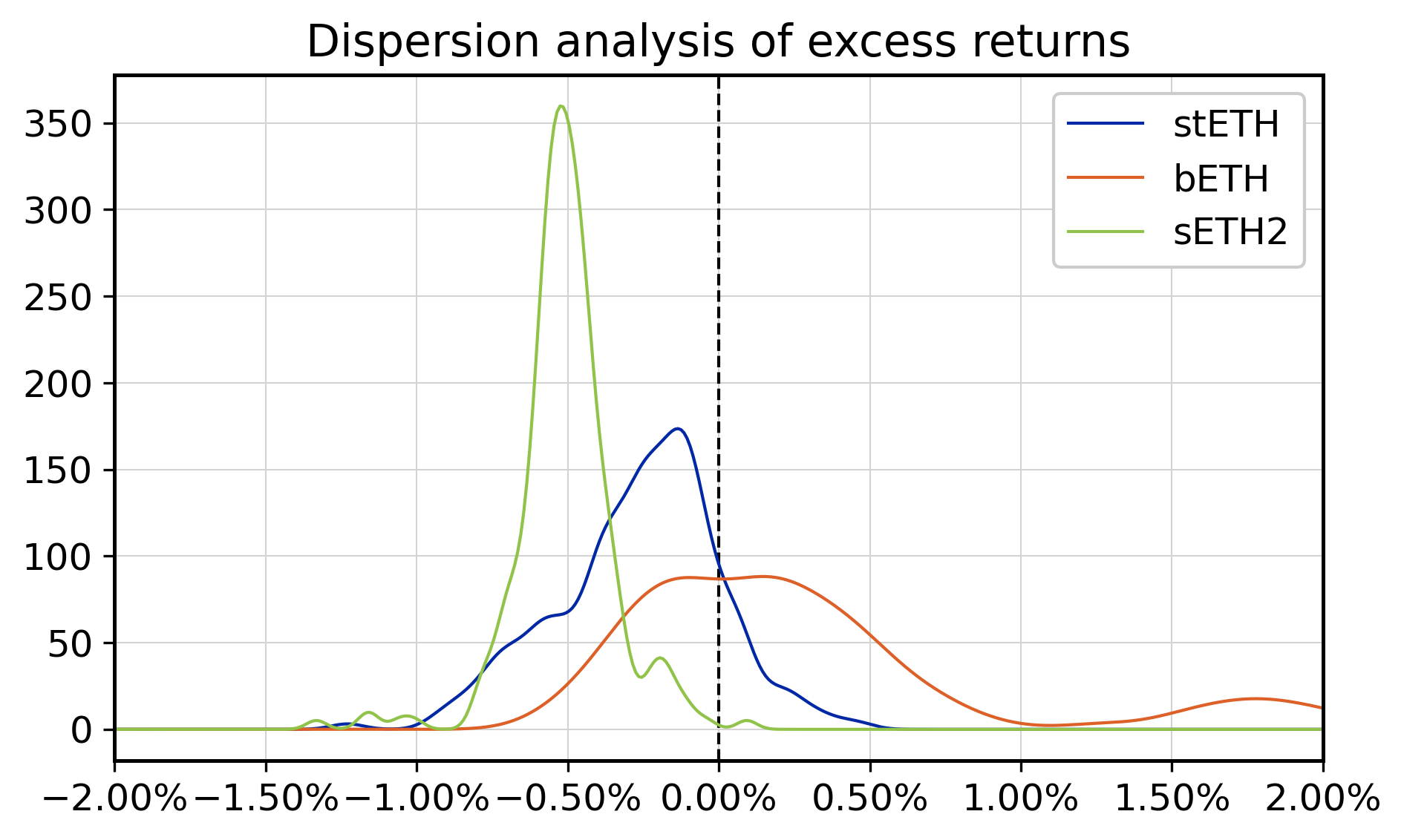} 
  \end{minipage} 
\hfill
    \begin{minipage}[c]{0.49\textwidth}
    \includegraphics[width=0.99\linewidth]{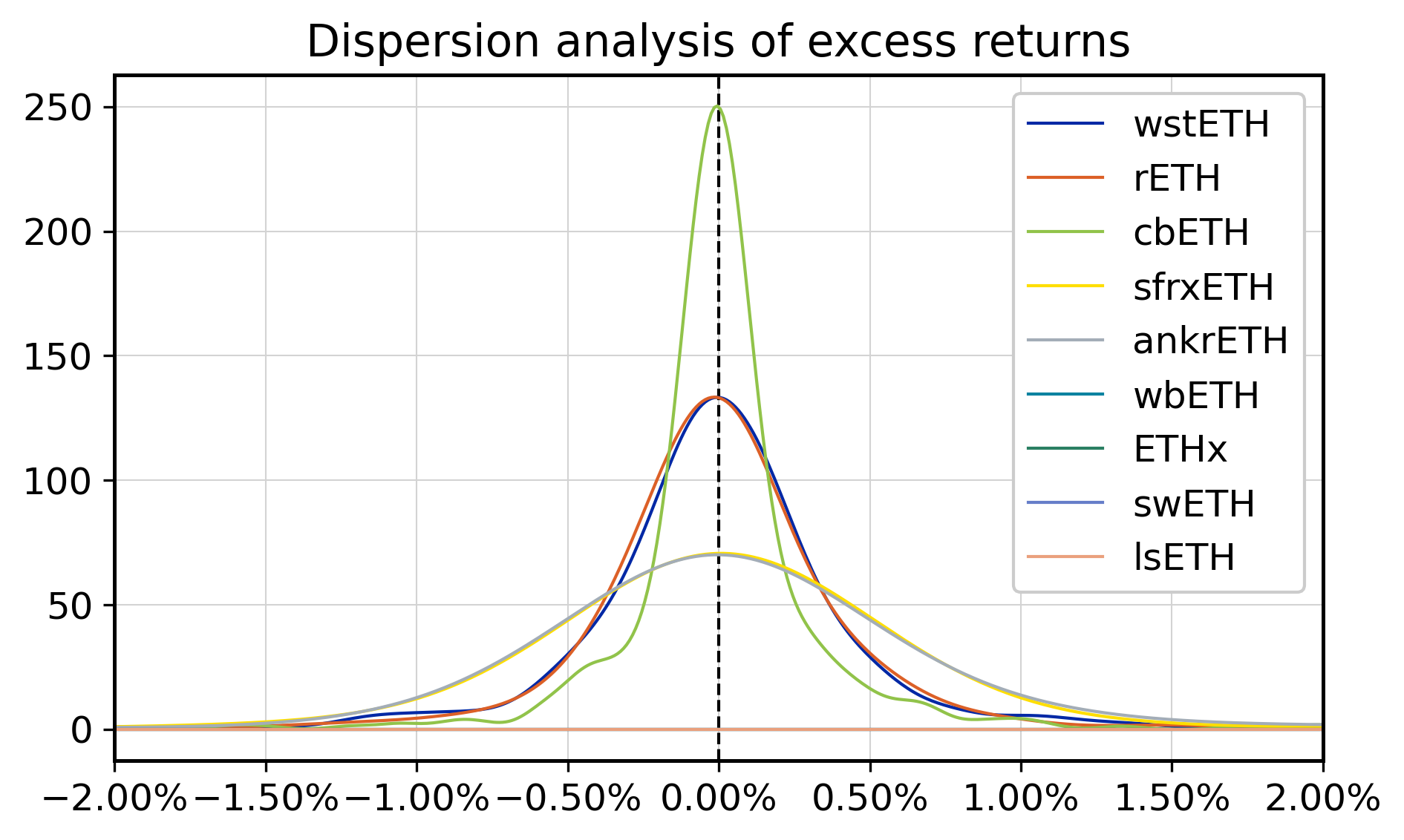} 
  \end{minipage} 
  \begin{minipage}[c]{0.49\textwidth}
    \includegraphics[width=.99\linewidth]{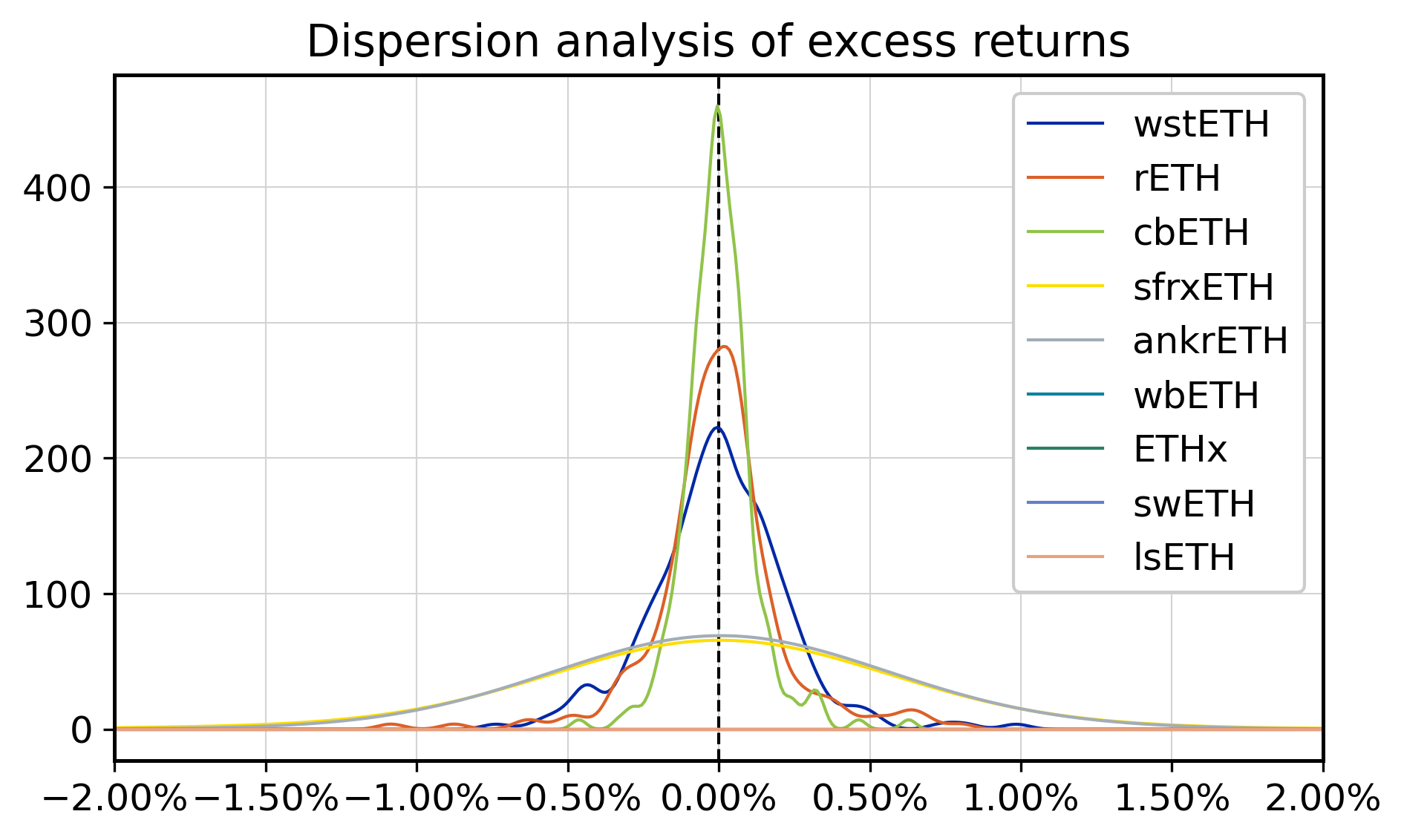} 
  \end{minipage} 
\caption{Dispersion analysis of rebase LSTs (top row) and reward-LSTs (bottom row), comparing data since token inception (left column) and after the Shanghai upgrade (right column). For rebase LSTs, the analysis measures the difference between the LST market price and the target value of 1 ETH. For reward-LSTs, it evaluates the difference between daily LST returns and the staking rate.}

  \label{fig:d_Ethereum}
\end{figure*}

The dispersion analysis of ETH LSTs is presented in Figure \ref{fig:d_Ethereum}. This analysis highlights: 
(i) The distribution of daily return differences between reward-LSTs and daily staking rewards.
ii) The deviation from the target value of 1 ETH for rebase-LSTs.  

The analysis of Figure \ref{fig:d_Ethereum} indicates that fees charged by LSPs introduce a negative shift in the distribution, while additional MEV rewards shift the distribution rightward. As a result, the overall returns for LSPs are slightly negative, with a negative median return. Among LSTs, cbETH - centralized LST issued by Coinbase - exhibits the lowest tracking error, whereas ankrETH demonstrates the highest deviation. After the Shanghai upgrade, which enabled ETH unstaking, deviations from the staking rate significantly decreased.



\paragraph*{Impact of Terra and FTX Collapses}
The crypto market faced severe disruptions in 2022 due to adverse conditions. The collapse of the algorithmic stablecoin UST on May 7, 2022, led to the Terra blockchain’s downfall, causing widespread market turmoil. Later, on November 2, 2022, the centralized exchange FTX became insolvent, freezing customers' digital assets.

During the period between the Terra collapse and FTX insolvency, both \textit{stETH} and \textit{rETH} underperformed compared to traditional staking. However, following the FTX collapse, \textit{rETH} outperformed traditional staking as investors preferred decentralized protocols over centralized ones. Rocket Pool, known for its permissionless validator selection, is considered more decentralized than Lido, which relies on a whitelist approved through DAO voting.

\paragraph*{Peg Analysis}

\begin{figure}[ht]
\centerline{
\includegraphics[width=1.05\linewidth]{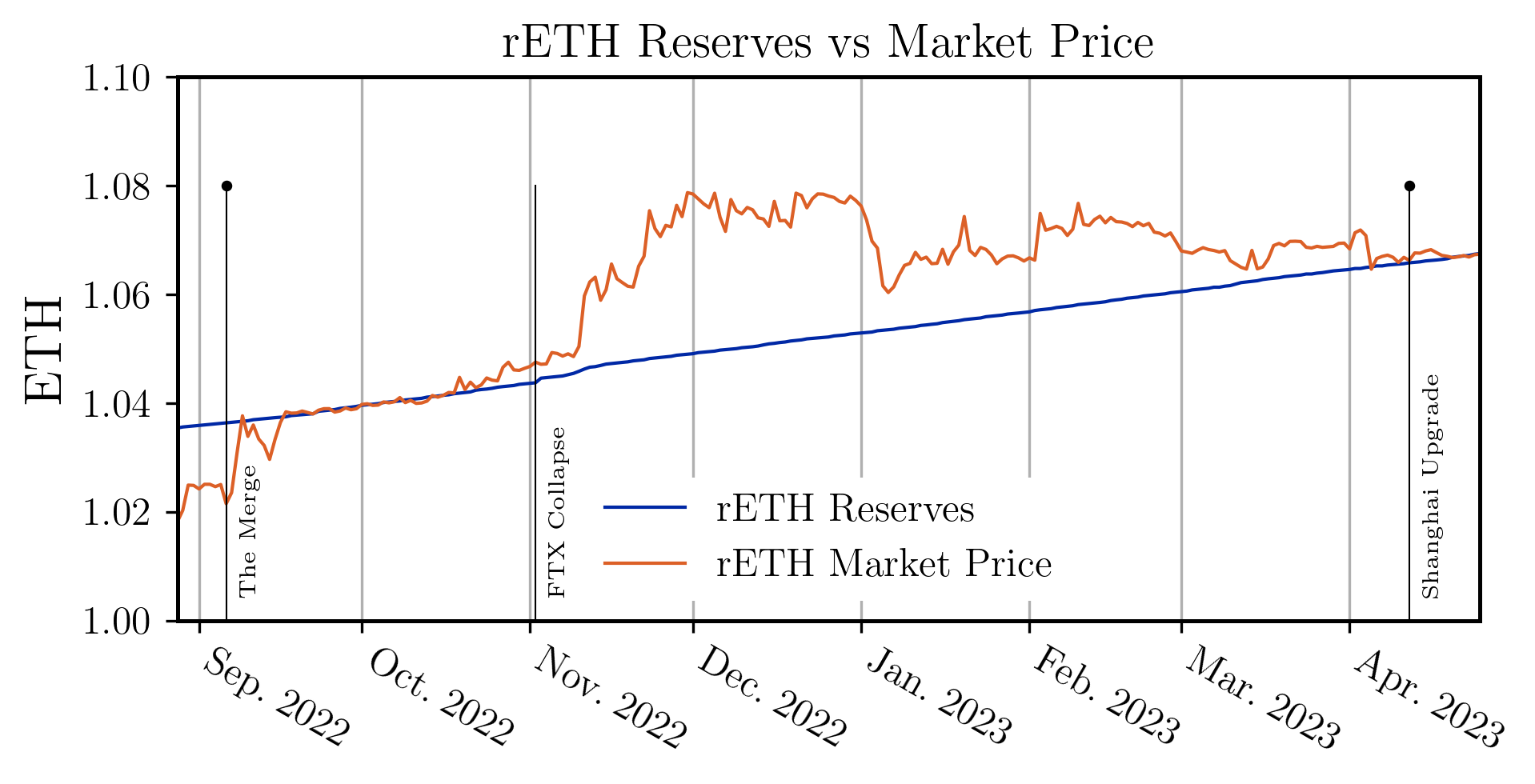}}
\caption{Divergence of the rETH market price from its reserve value during the period between the FTX insolvency and the Shanghai upgrade.}
\label{fig:peg_rETH}
\end{figure}

Prior to the Shanghai upgrade, the inability to unstake ETH caused deviations between market and fair (peg) values. Figure~\ref{fig:peg_rETH} illustrates the market and peg values for \textit{rETH}, the LST of Rocket Pool. Between the Terra-Luna collapse on May 7, 2022, and the FTX insolvency on November 2, 2022, \textit{rETH} was undervalued relative to its peg. If unstaking had been possible, arbitrageurs could have equalized the market and peg values.  
After the FTX collapse and until January 2023, \textit{rETH} was overpriced. This created arbitrage opportunities, as outlined in Algorithm \ref{alg:arbitrageHigh}. Arbitrageurs could mint \textit{rETH} at the Rocket Pool protocol for the protocol price and sell it on DEXs, such as Uniswap, at a higher market price. This overpricing reflected market aversion to centralized protocols post-FTX.

Before the Atlas upgrade in early 2023, Rocket Pool required 16 ETH in collateral from node operators to set up a validator~\cite{2023RocketPool:Documentation}. This limitation on creating new validators contributed to the upward de-peg of \textit{rETH} prices in 2022 and 2023. 

%% file: sections/04Restaking.tex
\section{Restaking}
    \label{sec:restaking}

 While liquid staking allows for the transferability of staked ETH and its use in DeFi, restaking extends this concept by enabling the same staked ETH to secure multiple protocols simultaneously. The protocols secured by restaking include oracles, data availability (DA) layers, bridges, decentralized sequencers of layer-2 (L2) blockchains, among others. This section explores the mechanics of Ethereum restaking, Bitcoin staking, cross-chain security, and the emerging concept of Liquid Restaking Tokens (LRTs), analogously to LSTs designed for trading and DeFi integration.

\subsection{Ethereum Restaking}
Restaking enables users to utilize their staked ETH to secure both the Ethereum network and additional protocols. There are two primary types of Ethereum restaking: i) native restaking utilize the staked ETH, or other staked tokens of PoS blockchains, and ii) liquid restaking that utylize as collateral LSTs.

\paragraph*{Native Restaking} 
Native restaking, as pioneered by EigenLayer~\cite{2023EigenLayer:Whitepaper}, is available only to users who operate Ethereum validator nodes. This process is facilitated by smart contracts that manage staked assets under a validator’s node and provide crypto-economic security for restaking protocols. Validators opting for native restaking are required to install additional software for the restaking module and agree to restaking terms, including additional slashing conditions. These validators secure Actively Validated Services (AVSs) such as data availability layers, bridges, and oracles.

\paragraph*{Liquid Restaking}
Liquid restaking use LSTs, allowing users to stake their LSTs with a restaking validator and receive another token representing their stake. These tokens can then be further restaked on resatking protocols like EigenLayer. Liquid restaking simplifies participation in restaking for users who do not run validator nodes.

\subsection{Liquid Restaking Tokens}
\label{sec:liquidrestaking}

Liquid restaking protocols, such as Ether.Fi,  act as intermediaries, managing validator operations and AVS selection on behalf of users. These protocols issue \emph{Liquid Restaking Tokens} (LRTs), which accrue interest and can be freely traded or integrated into DeFi to earn additional yields. LRTs allow users to easily enter and exit restaking positions while increasing leverage by reinvesting into DeFi protocols. The platforms simplify the setup process and enhance accessibility for users who wish to participate in restaking without technical expertise. The architecture, token modesl, and peg mechanism of LRTs protocols is the same as that of LSTs presented in the previous sections.

\subsection{Cross-Chain Security}
Cross-chain security enhances the economic security of PoS chains by enabling the staking of remote crypto assets in addition to native tokens. For example, cross-chain staking allows assets like BTC or staked ETH to secure other PoS chains.

Cross-chain security, also called \emph{mesh security}, was first proposed for the Cosmos ecosystem~\cite{2022MeshSecurity,2023MeshSecurity}, in which tokens staked on one PoS chain (the provider chain) are restaked to secure another PoS chain (the consumer chain). Mesh security allows the consumer chain to avoid high inflation rates of the native tokens - the approach that was used historically in order to attract stakers of their native tokens.


\subsection{Bitcoin Staking}
Bitcoin staking is a form for cross-chain security, in which native BTC tokens are used to secure PoS chains. 
In this model, native BTC holders lock their tokens on the Bitcoin network using a Bitcoin Staking script for a predetermined period, gaining voting power in a consumer PoS protocol. This process ensures that BTC stakers earn PoS staking rewards. Finality providers, acting as delegates, use the voting power of staked BTC to secure the PoS chain. If a finality provider attacks the system, the associated BTC is slashed, discouraging malicious behavior.

Whereas Bitcoin staking can be achieved by using bridged BTC tokens, remote Bitcoin staking, pioneered by the Babylon Protocol~\cite{Babylon2023BitcoinStaking}, does not require additional trust assumptions. BTC remains locked in a contract on the Bitcoin chain, with protocol violations reported back for slashing. This approach, while trustless, is limited by Bitcoin’s lack of a Turing-complete smart contract layer, requiring new solutions using Bitcoin’s scripting language.

%% file: sections/06LRT_Protocols.tex
\subsection{Comparison of Restaking Protocols}  
Restaking and liquid restaking represent emerging advancements in the DeFi space, garnering significant user attention and capital. Collectively, restaking protocols boast a TVL of over \$25.2 billion, with the majority concentrated in three major protocols: EigenLayer (\$15.07 billion), Babylon Protocol (\$5.5 billion), and Symbiotic (\$2.2 billion)\cite{2022DeFiCategories}. Notably, EigenLayer is the third-largest DeFi protocol by TVL. A brief overview of these protocols is given in Table \ref{tab:RestakingProtocolTVL}.

\begin{table*}[t]
\centering
\caption{Overview of major restaking protocols, highlighting total value locked (TVL), underlying assets, Actively Validated Services (AVS), Governance (Gov.) Token and associated metrics.}
\label{tab:RestakingProtocolTVL}
\begin{tabular}{|l|l|r|l|l|p{1.7cm}|r|}
\hline
\textbf{Protocol} & \textbf{Reference} & \textbf{TVL (\$B)} & \textbf{Underlying Assets} & \textbf{Restaking Rewards} & \textbf{Gov. Token} & \textbf{AVS} \\ 
\hline
EigenLayer          & \cite{2023EigenLayer:Whitepaper} & 15.07 & Staked ETH, LSTs, EIGEN & EIGEN token & EIGEN & Oracles, DA Layers \\ 
Babylon Protocol    & \cite{Babylon2023BitcoinStaking}    & 5.52  & BTC             & Directly to BTC Wallet & - & PoS Chain Security \\ 
Symbiotic           & \cite{2024Symbiotics}  & 2.22  &  ERC-20s incl LSTs, wBTC & ERC-20 for each vault & - & Permissionless \\ 
\hline
\end{tabular}
\end{table*}

\paragraph*{EigenLayer}  
On EigenLayer~\cite{2023EigenLayer:Whitepaper}, stakers can lock withdrawal credentials for natively staked ETH in an Eigenpod or send LSTs, EIGEN tokens, and soon ERC-20 tokens to a strategy contract. In all cases, they receive shares representing their stake, which can be delegated to a node operator. Stakers must delegate their entire restaked balance to a single operator. Operators use these shares to secure AVSs by opting into the AVS middleware if they meet the AVS requirements. While EigenLayer initially vetted operators, it plans to allow permissionless operator registration in the future. However, specific AVSs may require an additional whitelisting process.
EigenLayer's governance token is EIGEN, which is also used to pay rewards from restaking to the stakers.

\paragraph*{Babylon Protocol}  
The Babylon Protocol~\cite{Babylon2023BitcoinStaking} introduces Bitcoin staking, enabling Bitcoin liquidity to secure PoS chains. By integrating Bitcoin into the staking ecosystem, Babylon eliminates the need for bridging tokens, ensuring slashable guarantees while maintaining high security and reducing systemic risks.
Currently, there are no rewards for staking in Babylon, as it is in a locking-only phase without an active PoS chain. There are no PoS staking rewards or participation incentives during this phase. Babylon’s governance token is yet to be launched.

\paragraph*{Symbiotic}  
Symbiotic~\cite{2024Symbiotics} is a generalized, permissionless protocol. It offers a framework for developers to build their own restaking implementation and operator set.
Symbiotic allows anyone to create a node operator set, define assets to be used as collateral, and establish the purpose of the staking, such as securing AVSs. Slashing and reward distribution mechanisms are customizable and permissioned. Stakers deposit tokens into a default collateral contract and receive asset-specific collateral tokens. These tokens can then be sent to various vaults, each with its own restaking setup.
Each vault manages accounting, deposits, withdrawals, delegation strategies, reward distribution, and slashing mechanisms. Networks define their own criteria for accepted collateral, operator selection, and slashing conditions.
Symbiotic mints ERC-20 tokens representing restaking assets for each configured setup. Its governance token has yet to be launched.

%% file: sections/08Discussion.tex
\section{Discussion}
    \label{sec:discussion}

With over 10 million ETH staked via Liquid Staking Tokens (LSTs), they have emerged as the preferred method of staking ETH, accounting for 37\% of all staked ETH. They represent the largest category within DeFi in terms of capital, with a total value locked (TVL) of \$27 billion. Lido, the first Liquid Staking Provider (LSP), dominates the market, holding 31\% of the total staked ETH \cite{2023DuneLSDDeposit}.

\paragraph{Security Concerns with Lido}  
Lido's dominance raises concerns about Ethereum's blockchain security. As its share of staked ETH approaches the economic security threshold of one-third, an attacker could potentially delay the network's finality. Although Lido employs 30 node operators, each managing approximately 1\% of the staked ETH, these operators rely on Lido's software. A hack targeting Lido's software could compromise over one-third of the staked ETH, enabling attacks such as delayed finality and temporary transaction censorship.

\paragraph{Decentralization Efforts by RocketPool}  
RocketPool has introduced measures to enhance decentralization among node operators, currently operating over 3,000 nodes. However, historically, RocketPool struggled to attract sufficient node operators to stake ETH from depositors, causing its LST market price to de-peg upwards. RocketPool requires node operators to deposit collateral of 16 ETH to 8 ETH to run a validator, compared to the standard 32 ETH, while Stader protocol lowers this threshold to 4 ETH. This reduced collateral requirement decreases Ethereum's economic security threshold to 11\% and 4.13\%, respectively.

\paragraph{Distributed Validator Technology (DVT)}  
Distributed Validator Technology (DVT) aims to mitigate risks associated with single points of failure. By distributing validator keys across multiple machines, DVT enhances the resilience of the staking infrastructure and complicates malicious attacks. However, DVT increases infrastructure complexity and requires well-designed policies for distributing key shares among node operators. Improper distribution policies could reduce the attack threshold further, such as 2/3 of the signature threshold required for validator control. Simulations show that randomly distributing key shares decreases the likelihood of an attacker controlling the necessary threshold for validator compromise. 

\paragraph{Challenges in ETH Unstaking}  
Another challenge in Liquid Staking Provider (LSP) design is enforcing ETH unstaking and managing unstaking queues. While centralized protocols can control withdrawal keys, some decentralized protocols lack mechanisms to compel node validators to unstake ETH. Prolonged unstaking queues can exacerbate LST market de-pegs, especially when used as collateral in DeFi lending. Leveraged staking strategies magnify this risk, as significant de-pegs above liquidation thresholds could trigger cascading liquidations, further reducing token value. This scenario might force LSPs to unstake large amounts of ETH, potentially undermining Ethereum's blockchain security.

\paragraph{Emergence of Index Protocols and Institutional LSPs}  
To mitigate single LST de-pegging risks, index protocols like Asymmetry Finance \cite{2023AsymmetryFinance} have emerged, offering diversified baskets of LSTs from multiple providers. Additionally, institutional liquid staking providers, such as Liquid Collective, cater to corporate investors, employing Know Your Customer (KYC) processes for both node operators and depositors.

\paragraph{Alternative Staking Mechanisms in DeFi}  
While this discussion focuses on staking as an integral part of Proof of Stake (PoS) blockchains, DeFi also includes other staking mechanisms. For example, some protocols allow users to stake governance or liquidity provider (LP) tokens, locking them for specific periods to accumulate additional rewards.

\paragraph{Liquid Staking vs Derivatives}
Liquid Staking Tokens (LSTs) are often mistakenly referred to as Liquid Staking Derivatives (LSDs), which is a misclassification. In traditional finance, the value of a derivative contract is determined by a central counterparty, which is not the case for liquid staking tokens. Liquid staking represents tokenized staked ETH, with each token fully backed by reserves of staked ETH. Therefore, liquid staking tokens are derivatives, neither synthetic tokens. Synthetic tokens are backed by assets different from the target value and rely on financial engineering to maintain their pegged value.
Similarly, liquid restaking tokens are neither derivatives nor synthetic tokens. They are pegged tokens with their value directly tied to the underlying staked asset.

\paragraph{Restaking}
Restaking, despite being a relatively new concept, has attracted significant user interest and capital, establishing itself as the fourth largest category in DeFi with \$25 billion TVL. EigenLayer, the third largest DeFi protocol (following Lido, the largest), exemplifies the rapid growth and adoption of restaking, with \$15 billion controlled by this first restaking protocol. While the prospect of earning additional yield on top of staking is appealing, it raises questions about whether the associated risks are adequately compensated. Each new actively validated service (AVS) introduces additional layers of risk to the ecosystem, necessitating further research into the security and economic implications of restaking. Furthermore, security attacks on Ethereum PoS become more attractive as attackers might also target AVS secured by staked ETH. The concentration of restaking TVL in one protocol is another concern, as well as the payment of restaking rewards in governance tokens - EIGEN, which may introduce systemic vulnerabilities. 
Babylon Protocol, the second-largest restaking protocol with over \$5 billion in TVL, focuses on developing staking solutions for Bitcoin. While it enables voting power in other PoS networks, staking reward distributions is not yet operational and numerous technical and  remain unresolved, as highlighted in its whitepapers. 


\section{Related Work}
    \label{sec:relatedwork}

Liquid staking and its risk have not been vastly studied in the literature. Scharnowski et al.~\cite{Scharnowski2022LiquidDiscovery} conducted the first economic analysis of LSTs, focusing on the price disparity between LSTs and their corresponding native tokens.  The work~\cite{Gogol2023EmpiricalProtocols} differentiated three staking distribution models - reward, rebase, and dual and observed discrepancies between the protocol and market prices for LSTs on Ethereum before the Merge. The DeFi SoK~\cite{Gogol2023SoK:Risks} classified LSTs as pegged tokens with related de-peg risks. We detailed this classification with the technical mechanisms specific to LSTs and not present in design of other pegged tokens such as stablecoins. 

Building upon previous studies, this research established detailed LST taxonomy that includes key technical aspects such as the validators' management mechanisms that impact LST price trajectory and might lead to the price disparities. Our study further examined and discussed the technical limitations of LST protocols impacting the LST performance and risk, especially the resistance to de-peg in the aftermath of extreme market events. Only by analyzing the token designs and implementation limitations can it be explained under which conditions the discrepancies between market and protocol prices lead to arbitrage opportunities. Other related work focused on the specific area within the LST application, which we briefly present further. 

\paragraph*{Principal Agent Problem in Liquid Staking Protocols}

Although liquid staking allows stakers to earn rewards while retaining liquidity, it creates new types of risks - a principal–agent problem~\cite{tzinas2023principal}: stakers (principals) cede control to liquid staking protocols (agents), who decide how to delegate stake among validators. Misaligned incentives can lead protocols to prioritize profit or convenience over security, potentially increasing centralization or slashing risk. Possible mitigation strategies include coverage against slashing, transparent validator performance metrics, decentralized governance, and performance-based fees.

\paragraph*{Liquid Staking Tokens in DeFi}
Liquid staking is widely used in DeFi, predominantly in liquidity provisions in AMMs and lending protocols. According to empirical research\cite{gogol2024LSTinAMM}, the liquidity provision of LSTs to AMMs is less profitable than staking, leading to the loss called "loss-versus-staking". While trading fees in AMMs often compensate liquidity providers (LPs) for impermanent loss, staking of entire wealth is still more profitable that rewards from liquidity provisions~\cite{gogol2024LSTinAMM}. 

The corresponding results were observed in the empirical studies~\cite{xiong2024ExpoLSD}, which also noted and analyzed the price disparities of rETH protocol and market prices. Without investigating the design of RocketPool protocol, 
the disparities were assigned to the staking queues at RocketPool. In practice the arbitrage opportunities on rETH could be exploited but arbitrageurs would need to operate a new validators with a collateral of 16ETH, which collateral requirement by RocketPool to operate validator before their Atlas upgrade. This instance highlights the critical importance of LSP protocol designs, which is presented in this SoK.

Other studies in DeFi explored the implications of leveraging LSTs within DeFi lending protocols~\cite{xiong2024leverageLST}. Their simulation results suggest that leveraged staking can amplify the risk of cascading liquidations by introducing heightened selling pressure from liquidations, posing systemic risks to the broader ecosystem. However, empirical data showed that the majority of leverage staking positions achieved higher returns than staking~\cite{xiong2024leverageLST}.

\paragraph*{Restaking Risks}
At present, the risks associated with restaking are not comprehensively analyzed. While restaking offer an opportunity to horizontally multiply the stake by securing other actively validated services (AVS) it also multiplies the slashing risks as the stake is locked in multiple services. The implicit leverage exerted on the restaked stake may also have implications on the stability and volatility of restaking tokens in DeFi such as liquid restaking tokens. In their analysis of validator reuse across multiple services in a restaking protocol, Druvasula and Roughgarden~\cite{durvasula2024robust} characterize robust security based on the buffer between attack costs and profits. They derive explicit bounds on worst-case stake loss, propose overcollateralization conditions (including local analogs), and suggest polynomial-time computable criteria that help participants prevent cascading attacks. Chitra and Pai ~\cite{chitra2024much} investigate how restaking protocols can minimize the security needed for restaked services. They extend the model of ~\cite{durvasula2024robust}, incorporating token-based incentives, strategic adversaries, and rational users to show that properly designed incentives help avoid large-scale cascading failures.Their findings suggest that restaking protocols can remain secure if incentive mechanisms are managed effectively.

%% file: sections/09Conclusions.tex
\section{Conclusions}
    \label{sec:conclusions}

This paper provides a systematic study of liquid staking protocols and tokens. It further presents the emerging solutions in restaking such as Ethereum restaking, Bitcoin staking and cross-chain security. 
A comprehensive taxonomy was established to analyze the mechanisms and economic models underlying liquid staking, including node validator selection, staking reward distribution, and token models. This work positioned Liquid Staking Tokens (LSTs) within the broader taxonomy of Decentralized Finance (DeFi) and pegged tokens,

Empirical evaluations were performed to assess the performance, security, and decentralization of LSTs. The study highlights the impact of protocol architecture on tracking staking rewards and maintaining peg stability, demonstrating that centralized protocols exhibit superior accuracy and scalability.
The empirical findings revealed that while LSTs generally track staking rewards effectively, temporary de-pegs occur during significant market events. Decentralized LSTs, like RocketPool's rETH, experience upward de-pegs when validator capacity is insufficient. Following Ethereum’s Shanghai upgrade, improvements in tracking staking rewards were observed across LSTs.

Lastly, this SoK provided a review of the emerging protocols, white-papers and literature summarizing current research on liquid staking and restaking. Restaking, although it has already attracted the interest and capital of the users, introduced new risks by increasing the load on the PoS consensus, and requires further research on its broader implications.

%% file: sections/98EthereumSecurity.tex
\subsection{LST Impact on Ethereum Security}
    \label{sec:ETHSecurity}

LSPs do not directly manage validators but establish a network of node operators responsible for validator operations. In the permissioned model, these node operators are whitelisted by the protocol, often through a DAO vote, effectively preventing the inclusion of malicious actors.
In the permissionless model, node operators are required to deposit collateral to operate a validator, with the remaining ETH provided by the staking pool of the LSP. Consequently, attackers on the Ethereum network no longer need to provide the full 32 ETH to operate a validator but only a portion covering the collateral requirements of the LSP. This reduction decreases the attack threshold, as shown in Table \ref{tab:LSDAttackThreshold}.

Further reductions in the attack threshold are possible with naive implementation of Distributed Validator Technology (DVT) without random key distribution. Assuming only 2/3 of the key shares are required to operate a validator and node operators can select key shares, attack thresholds may be further diminished by 2/3. However, LSPs can mitigate this risk by implementing random distribution of validator key shares among node operators.


Assuming that an LSP has $n$ node operators and $m$ validators, and each validator key is split into $k$ shares, there are $mk$ key shares distributed randomly among node operators. At least $\pi$ key shares are required to control a validator. Each node operator must provide collateral of $\frac{32\text{ ETH}}{\pi}$ to receive one key share. The probability that one node operator receives at least $\ell$ key shares for a single validator is:

\begin{align*}\label{eq:DVT}
    P(m, k, \ell) &= m \cdot \frac{k}{mk} \cdot \frac{k-1}{mk-1} \cdot \ldots \cdot \frac{k-\ell+1}{mk-\ell+1} \\
    &= \frac{k!}{(mk)!} \cdot \frac{(mk-\ell)!}{(k-\ell)!}
\end{align*}

\begin{figure*}[tbh]
\centering
  \begin{minipage}[c]{0.4\textwidth}
    \includegraphics[width=0.99\linewidth]{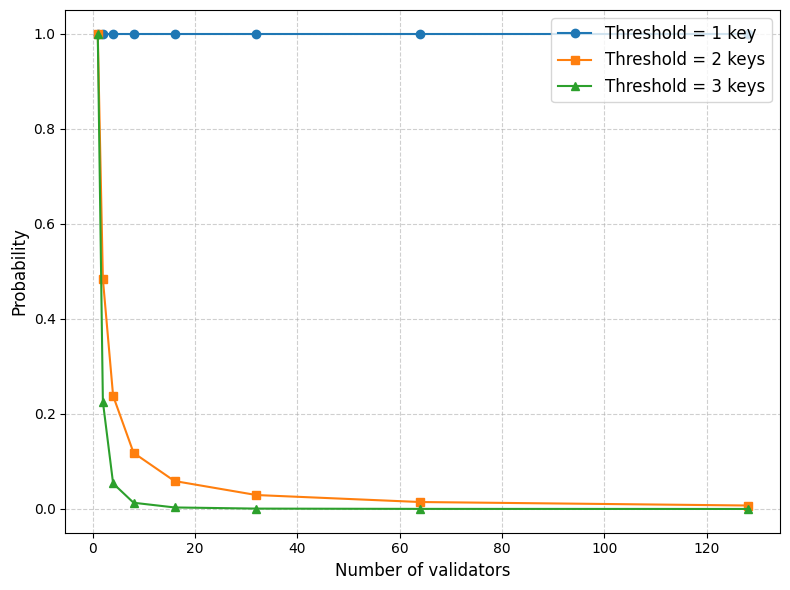}
  \end{minipage}
  \begin{minipage}[c]{0.4\textwidth}
    \includegraphics[width=.99\linewidth]{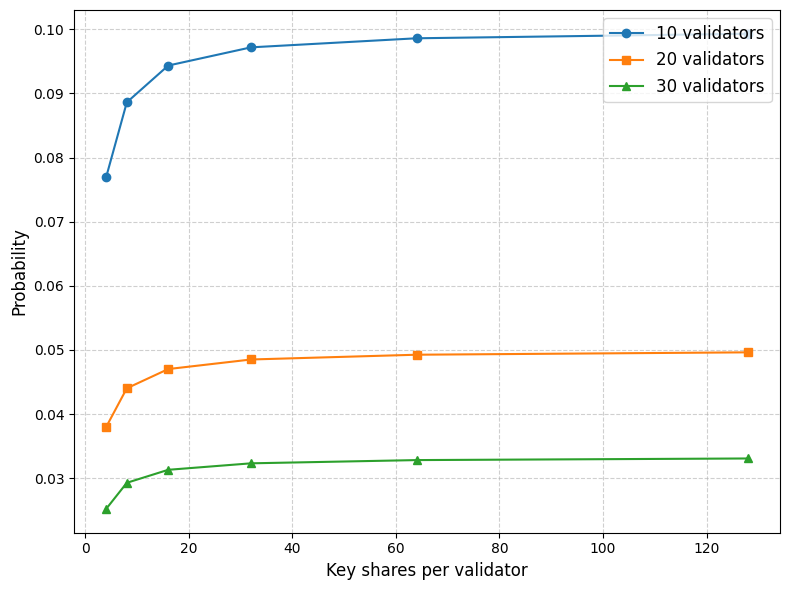}
  \end{minipage}

  \medskip

  \begin{minipage}[c]{0.4\textwidth}
    \includegraphics[width=.99\linewidth]{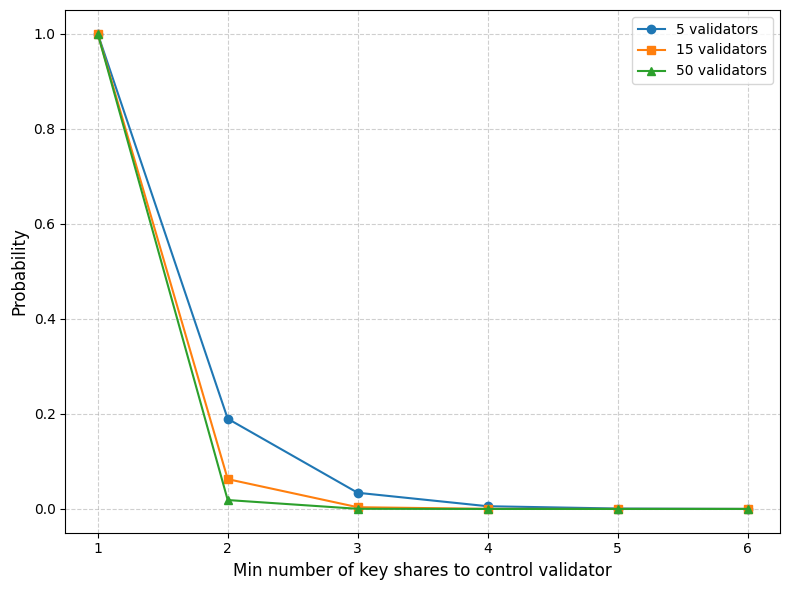}
  \end{minipage}
  \begin{minipage}[c]{0.4\textwidth}
    \includegraphics[width=.99\linewidth]{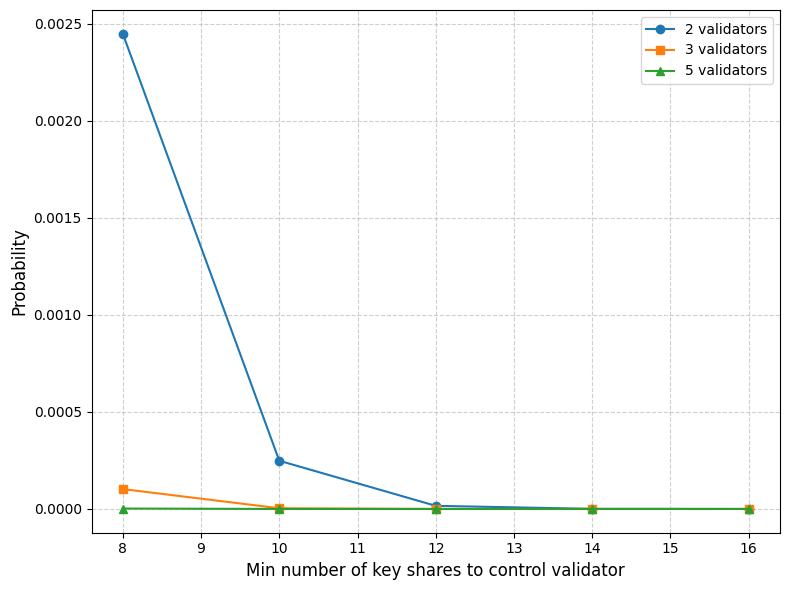}
  \end{minipage}
\caption{Probability of a single node operator controlling a validator based on the number of validators, key shares per validator, and the minimum signature threshold. Validator keys are divided into 16 key shares, with a 2-share signature threshold unless otherwise specified.}
  \label{fig:DVTCharts}
\end{figure*}

The probability that a single node operator controls $\pi$ key shares of one validator decreases significantly as a function of the signature threshold $\pi$ and the number of validators $m$, as shown in Figure \ref{fig:DVTCharts}.

